\documentclass{jcgt}

\usepackage{amsmath}
\usepackage{listings}
\usepackage{xcolor}
\usepackage{siunitx}
\usepackage{url}
\usepackage{subcaption}
\usepackage{algorithm}
\usepackage{algorithmicx}
\usepackage{algpseudocode}
\usepackage{mathtools}
\usepackage{xfrac}
\usepackage{comment}
\usepackage[font={small,it}]{caption}
\usepackage{tikz}
\usetikzlibrary{positioning}

\usepackage[export]{adjustbox}

\setciteauthor{Jamie Portsmouth, Matthias Raab, Laurent Belcour, and Francis Liu}
\setcitetitle{The Minimal Retroreflective Microfacet Model}

\accepted{2025-12-24}{2026-04-12}{2026-05-22}{Bernd Bickel}{15}{1}{60}{75}{2026}
\seturl{http://jcgt.org/published/0015/01/04/}

\newcommand{\vecv}{\mathbf{v}}
\newcommand{\vecvv}{\mathbf{v'}}
\newcommand{\vecl}{\mathbf{l}}
\newcommand{\vecll}{\mathbf{l'}}
\newcommand{\vecn}{\mathbf{n}}
\newcommand{\vecm}{\mathbf{m}}
\newcommand{\vech}{\mathbf{h}}
\newcommand{\vechr}{\mathbf{h}_r}
\newcommand{\vecht}{\mathbf{h}_t}
\newcommand{\vecb}{\mathbf{b}}
\newcommand{\vecbr}{\mathbf{b}_r}
\newcommand{\vecbt}{\mathbf{b}_t}
\newcommand{\vecx}{\mathbf{x}}

\newcommand{\vecy}{\mathbf{y}}

\newcommand{\etav}{\eta_{\mathbf{v}}}
\newcommand{\etal}{\eta_{\mathbf{l}}}

\newcommand{\vdot}[2]{#1 \cdot #2}
\newcommand{\avdot}[2]{\lvert\vdot{#1}{#2}\rvert}
\newcommand{\cldot}[2]{\langle #1 , #2 \rangle}

\newcommand{\Rn}{\mathbf{R_{\mathbf{n}}}}

\newcommand{\diff}{\, \mathrm{d}}

\includecomment{introduction}
\includecomment{existing-models}
\includecomment{eon}
\includecomment{eon-sampling}
\includecomment{conclusion}

\lstdefinestyle{snippet}{
  float=tp,
  floatplacement=tbp,
  abovecaptionskip=5pt
}

\lstdefinelanguage{GLSL}%
{%
	morekeywords={%
		false,FALSE,NULL,true,TRUE,%
		__LINE__,__FILE__,__VERSION__,GL_core_profile,GL_es_profile,GL_compatibility_profile,%
		precision,highp,mediump,lowp,%
		void,bool,int,uint,float,double,vec2,vec3,vec4,dvec2,dvec3,dvec4,bvec2,bvec3,bvec4,ivec2,ivec3,ivec4,uvec2,uvec3,uvec4,mat2,mat3,mat4,mat2x2,mat2x3,mat2x4,mat3x2,mat3x3,mat3x4,mat4x2,mat4x3,mat4x4,dmat2,dmat3,dmat4,dmat2x2,dmat2x3,dmat2x4,dmat3x2,dmat3x3,dmat3x4,dmat4x2,dmat4x3,dmat4x4,sampler1D,sampler2D,sampler3D,image1D,image2D,image3D,samplerCube,imageCube,sampler2DRect,image2DRect,sampler1DArray,sampler2DArray,image1DArray,image2DArray,samplerBuffer,imageBuffer,sampler2DMS,image2DMS,sampler2DMSArray,image2DMSArray,samplerCubeArray,imageCubeArray,sampler1DShadow,sampler2DShadow,sampler2DRectShadow,sampler1DArrayShadow,sampler2DArrayShadow,samplerCubeShadow,samplerCubeArrayShadow,isampler1D,isampler2D,isampler3D,iimage1D,iimage2D,iimage3D,isamplerCube,iimageCube,isampler2DRect,iimage2DRect,isampler1DArray,isampler2DArray,iimage1DArray,iimage2DArray,isamplerBuffer,iimageBuffer,isampler2DMS,iimage2DMS,isampler2DMSArray,iimage2DMSArray,isamplerCubeArray,iimageCubeArray,atomic_uint,usampler1D,usampler2D,usampler3D,uimage1D,uimage2D,uimage3D,usamplerCube,uimageCube,usampler2DRect,uimage2DRect,usampler1DArray,usampler2DArray,uimage1DArray,uimage2DArray,usamplerBuffer,uimageBuffer,usampler2DMS,uimage2DMS,usampler2DMSArray,uimage2DMSArray,usamplerCubeArray,uimageCubeArray,struct,%
		gl_BackColor,gl_BackLightModelProduct,gl_BackLightProduct,gl_BackMaterial,gl_BackSecondaryColor,gl_ClipDistance,gl_ClipPlane,gl_ClipVertex,gl_Color,gl_DepthRange,gl_DepthRangeParameters,gl_EyePlaneQ,gl_EyePlaneR,gl_EyePlaneS,gl_EyePlaneT,gl_Fog,gl_FogCoord,gl_FogFragCoord,gl_FogParameters,gl_FragColor,gl_FragCoord,gl_FragData,gl_FragDepth,gl_FrontColor,gl_FrontFacing,gl_FrontLightModelProduct,gl_FrontLightProduct,gl_FrontMaterial,gl_FrontSecondaryColor,gl_InstanceID,gl_Layer,gl_LightModel,gl_LightModelParameters,gl_LightModelProducts,gl_LightProducts,gl_LightSource,gl_LightSourceParameters,gl_MaterialParameters,gl_ModelViewMatrix,gl_ModelViewMatrixInverse,gl_ModelViewMatrixInverseTranspose,gl_ModelViewMatrixTranspose,gl_ModelViewProjectionMatrix,gl_ModelViewProjectionMatrixInverse,gl_ModelViewProjectionMatrixInverseTranspose,gl_ModelViewProjectionMatrixTranspose,gl_MultiTexCoord0,gl_MultiTexCoord1,gl_MultiTexCoord2,gl_MultiTexCoord3,gl_MultiTexCoord4,gl_MultiTexCoord5,gl_MultiTexCoord6,gl_MultiTexCoord7,gl_Normal,gl_NormalMatrix,gl_NormalScale,gl_ObjectPlaneQ,gl_ObjectPlaneR,gl_ObjectPlaneS,gl_ObjectPlaneT,gl_Point,gl_PointCoord,gl_PointParameters,gl_PointSize,gl_Position,gl_PrimitiveIDIn,gl_ProjectionMatrix,gl_ProjectionMatrixInverse,gl_ProjectionMatrixInverseTranspose,gl_ProjectionMatrixTranspose,gl_SecondaryColor,gl_TexCoord,gl_TextureEnvColor,gl_TextureMatrix,gl_TextureMatrixInverse,gl_TextureMatrixInverseTranspose,gl_TextureMatrixTranspose,gl_Vertex,gl_VertexID,%
		gl_MaxClipPlanes,gl_MaxCombinedTextureImageUnits,gl_MaxDrawBuffers,gl_MaxFragmentUniformComponents,gl_MaxLights,gl_MaxTextureCoords,gl_MaxTextureImageUnits,gl_MaxTextureUnits,gl_MaxVaryingFloats,gl_MaxVertexAttribs,gl_MaxVertexTextureImageUnits,gl_MaxVertexUniformComponents,%
		abs,acos,all,any,asin,atan,ceil,clamp,cos,cross,degrees,dFdx,dFdy,distance,dot,equal,exp,exp2,faceforward,floor,fract,ftransform,fwidth,greaterThan,greaterThanEqual,inversesqrt,length,lessThan,lessThanEqual,log,log2,matrixCompMult,max,min,mix,mod,noise1,noise2,noise3,noise4,normalize,not,notEqual,outerProduct,pow,radians,reflect,refract,shadow1D,shadow1DLod,shadow1DProj,shadow1DProjLod,shadow2D,shadow2DLod,shadow2DProj,shadow2DProjLod,sign,sin,smoothstep,sqrt,step,tan,texture1D,texture1DLod,texture1DProj,texture1DProjLod,texture2D,texture2DLod,texture2DProj,texture2DProjLod,texture3D,texture3DLod,texture3DProj,texture3DProjLod,textureCube,textureCubeLod,transpose,%
		rgb
	},
	sensitive=true,%
	morecomment=[s]{/*}{*/},%
	morecomment=[l]//,%
	morestring=[b]",%
	morestring=[b]',%
	moredelim=*[directive]\#,%
	moredirectives={define,defined,elif,else,if,ifdef,endif,line,error,ifndef,include,pragma,undef,warning,extension,version},%
  classoffset=1, 
  keywords={retroGGX, break,case,continue,default,discard,do,else,for,if,return,switch,while,define},%
  keywordstyle=\color{codeblue}\bfseries,%
  classoffset=0%
  }[keywords,comments,strings,directives]%

\definecolor{backcolour}{rgb}{1, 1, 1}
\definecolor{codegreen}{rgb}{0,0.6,0}
\definecolor{codegray}{rgb}{0.5,0.5,0.5}
\definecolor{codepurple}{rgb}{0.58,0,0.82}
\definecolor{codeblue}{rgb}{0,0.3,0.6}

\begin{document}

\title{The Minimal Retroreflective Microfacet Model}

\author{\begin{tabular}{@{}c@{\hspace{1.5em}}c@{\hspace{1.5em}}c@{\hspace{1.5em}}c@{}}
  Jamie Portsmouth~\href{https://orcid.org/0000-0003-4261-7730}{\includegraphics[width=8pt]{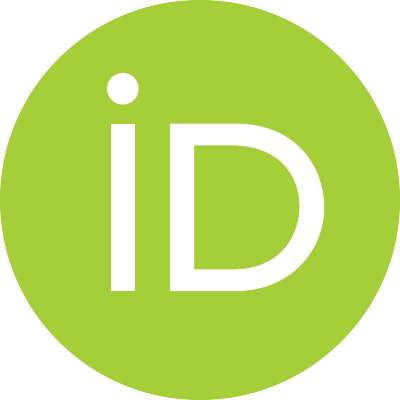}} &
  Matthias Raab~\href{https://orcid.org/0009-0006-7549-6727}{\includegraphics[width=8pt]{ORCIDlogo}} &
  Laurent Belcour~\href{https://orcid.org/0000-0002-1982-0717}{\includegraphics[width=8pt]{ORCIDlogo}} &
  Francis Liu~\href{https://orcid.org/0009-0009-2922-5784}{\includegraphics[width=8pt]{ORCIDlogo}} \\
  Autodesk & NVIDIA & Intel & NVIDIA
\end{tabular}}

\teaser{
    \begin{tikzpicture}[baseline=(img.south)]
      \node[anchor=south west, inner sep=0] (img) {\includegraphics[width=0.31\linewidth]{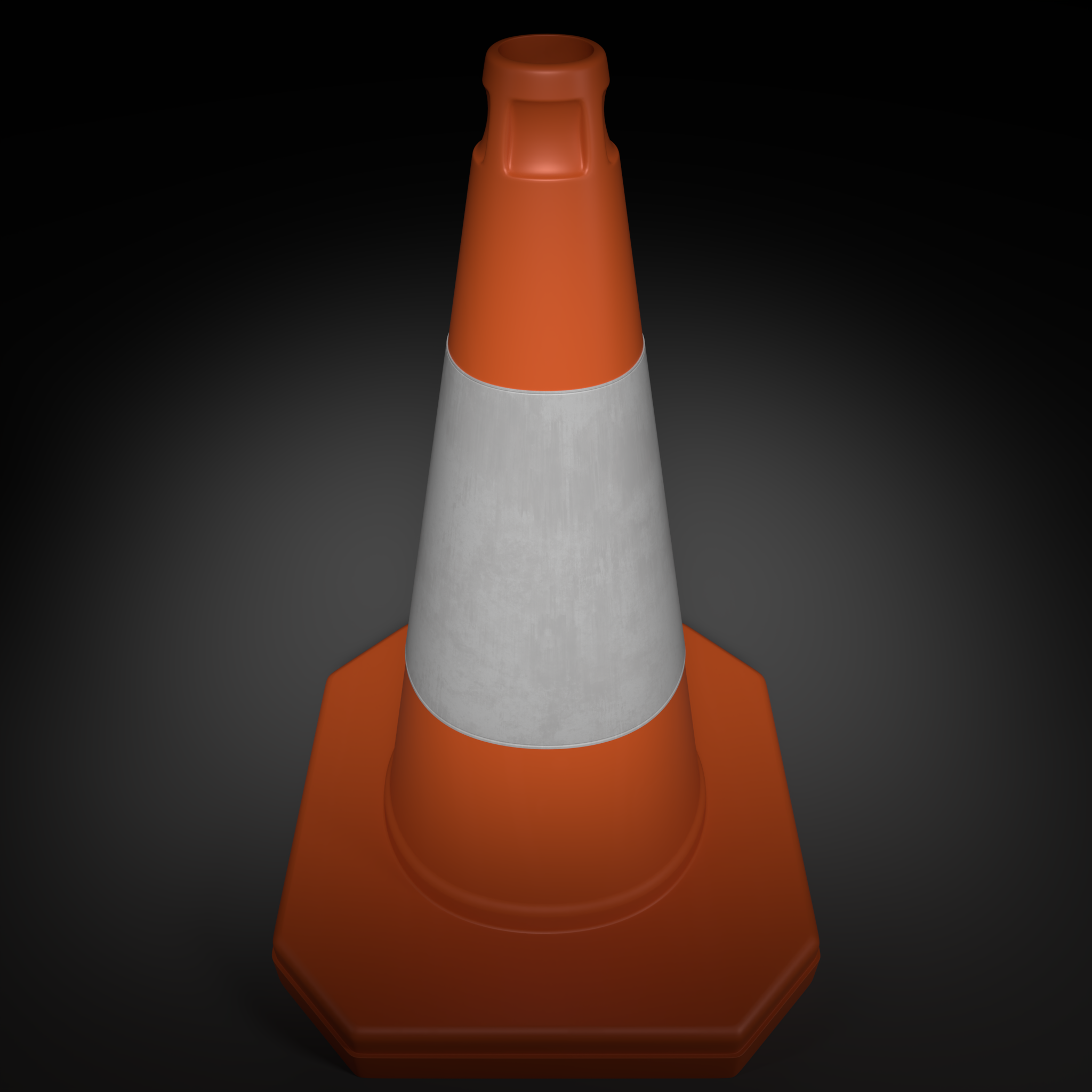}};
      \node[anchor=north west, text=white, font=\sffamily\bfseries\footnotesize, inner sep=3pt] at (img.north west) {GGX};
    \end{tikzpicture}%
    \begin{tikzpicture}[baseline=(img.south)]
      \node[anchor=south west, inner sep=0] (img) {\includegraphics[width=0.31\linewidth]{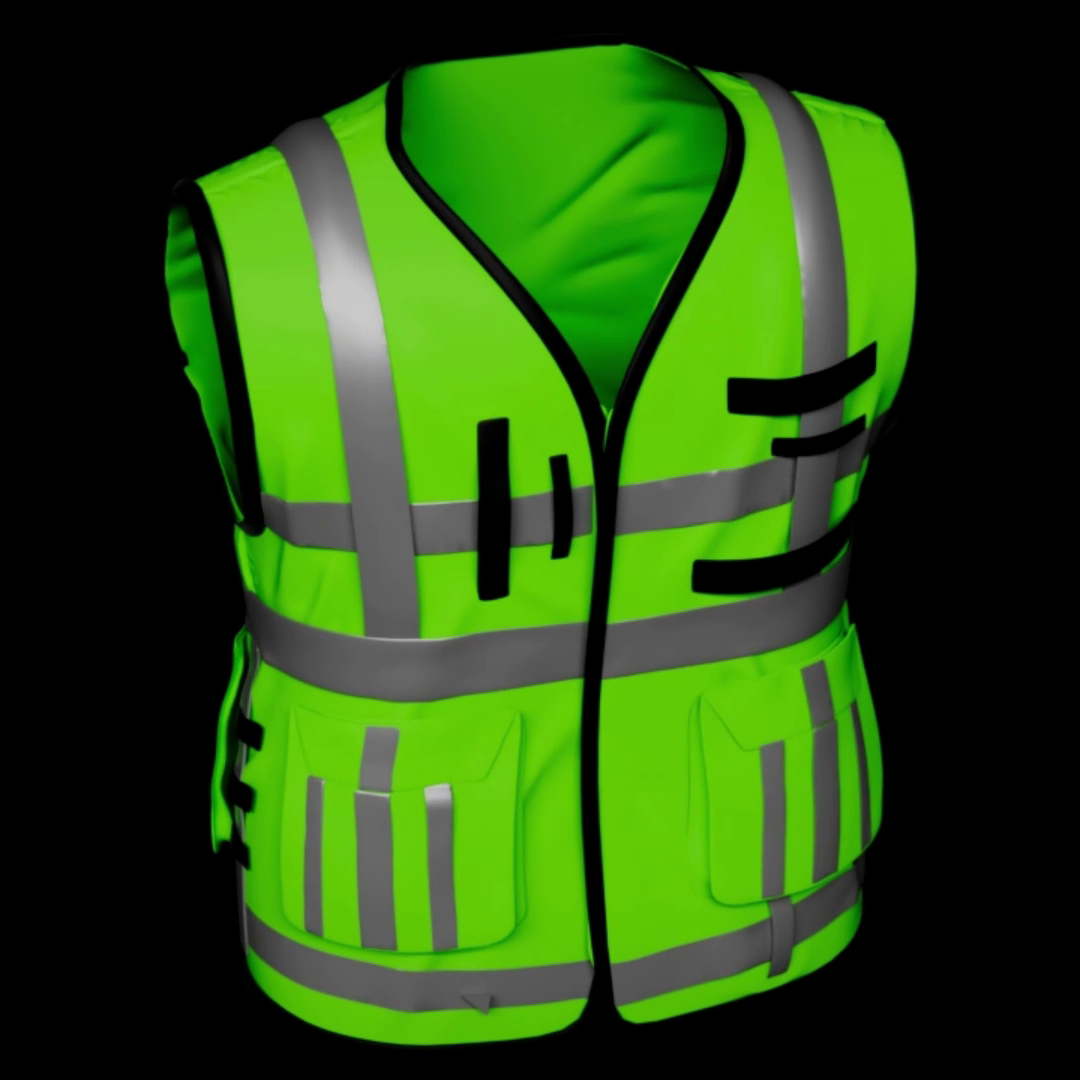}};
      \node[anchor=north west, text=white, font=\sffamily\bfseries\footnotesize, inner sep=3pt] at (img.north west) {GGX};
    \end{tikzpicture}%
    \begin{tikzpicture}[baseline=(img.south)]
      \node[anchor=south west, inner sep=0] (img) {\includegraphics[width=0.31\linewidth]{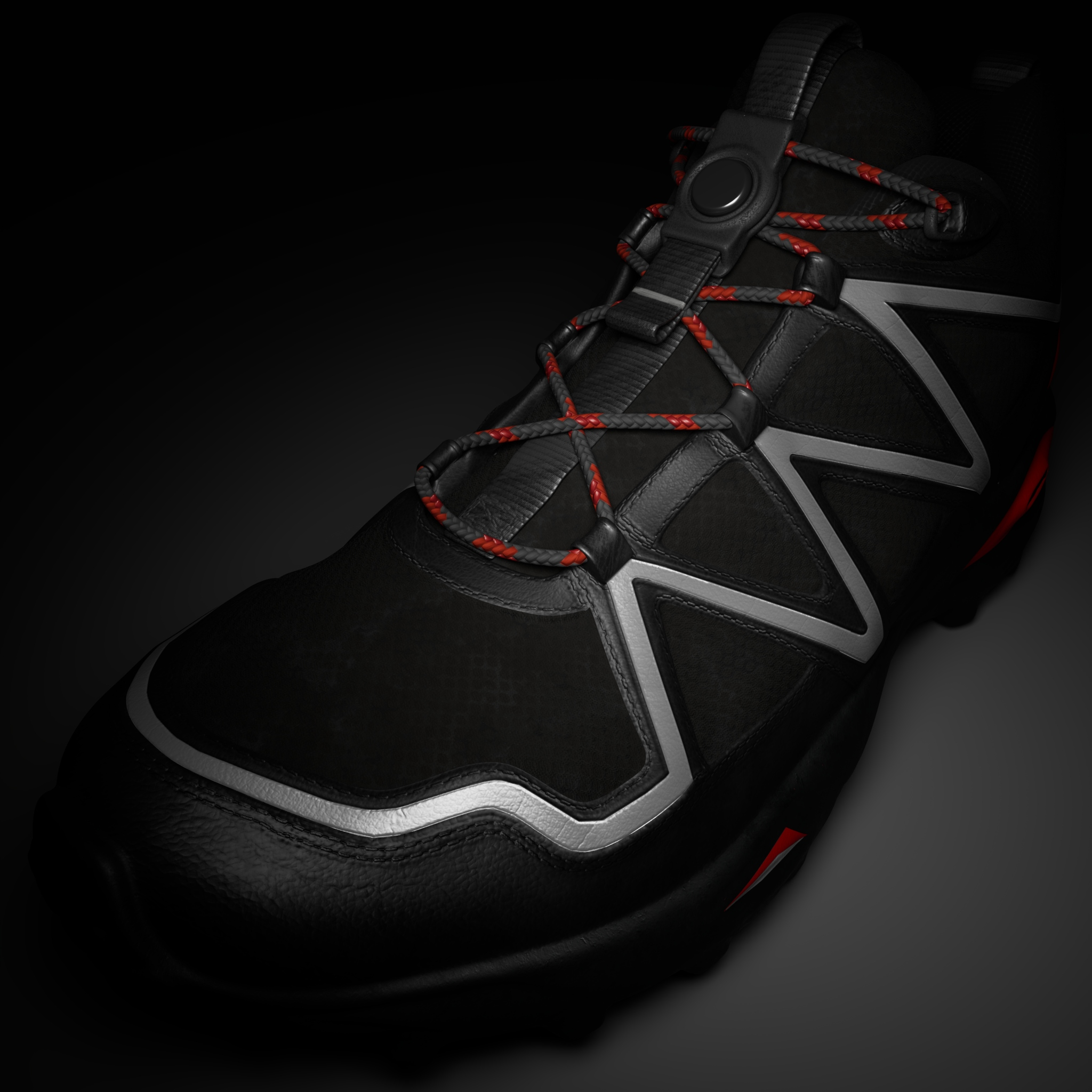}};
      \node[anchor=north west, text=white, font=\sffamily\bfseries\footnotesize, inner sep=3pt] at (img.north west) {GGX};
    \end{tikzpicture} \\
    \vspace{0.06cm}
    \begin{tikzpicture}[baseline=(img.south)]
      \node[anchor=south west, inner sep=0] (img) {\includegraphics[width=0.31\linewidth]{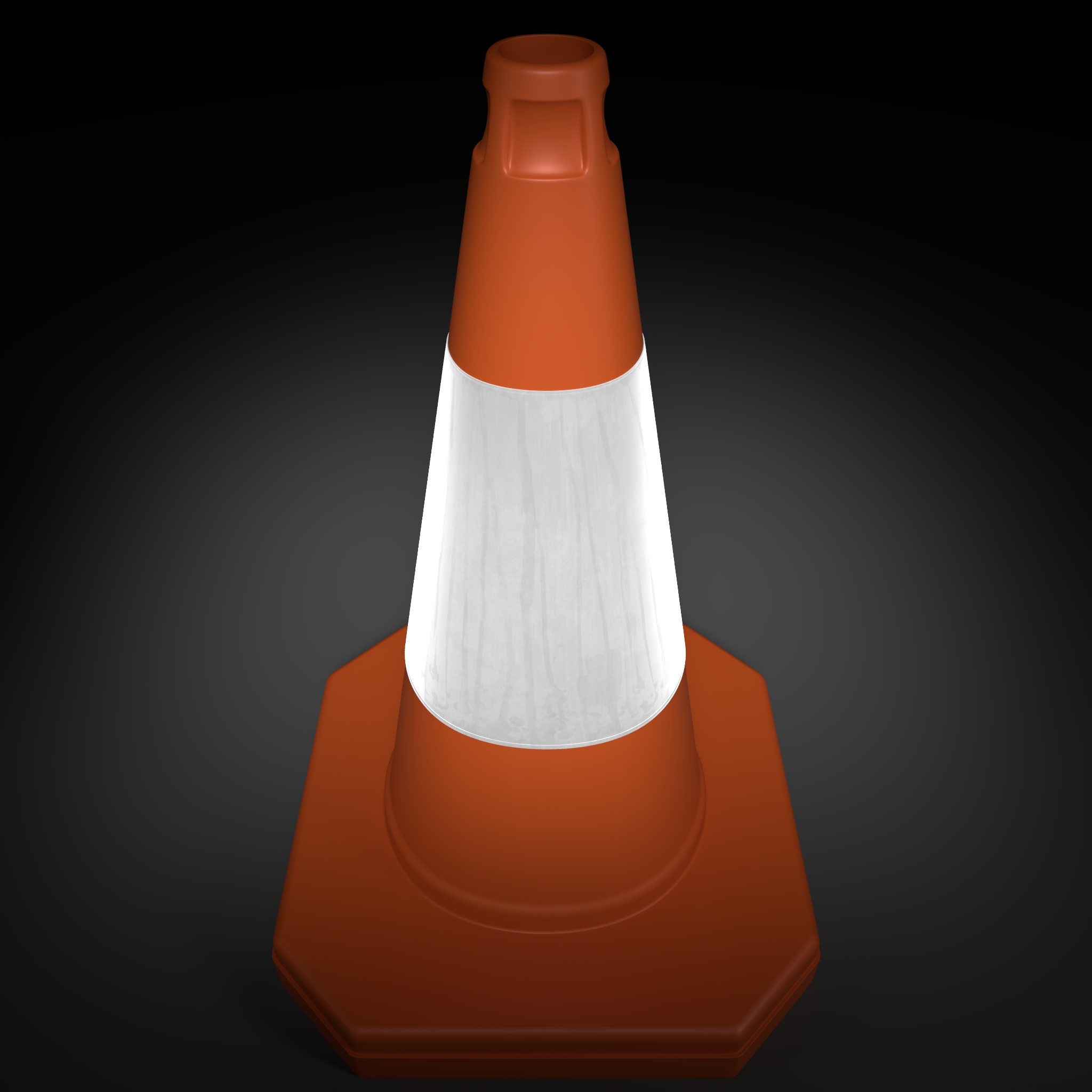}};
      \node[anchor=north west, text=white, font=\sffamily\bfseries\footnotesize, inner sep=3pt] at (img.north west) {MRM};
    \end{tikzpicture}%
    \begin{tikzpicture}[baseline=(img.south)]
      \node[anchor=south west, inner sep=0] (img) {\includegraphics[width=0.31\linewidth]{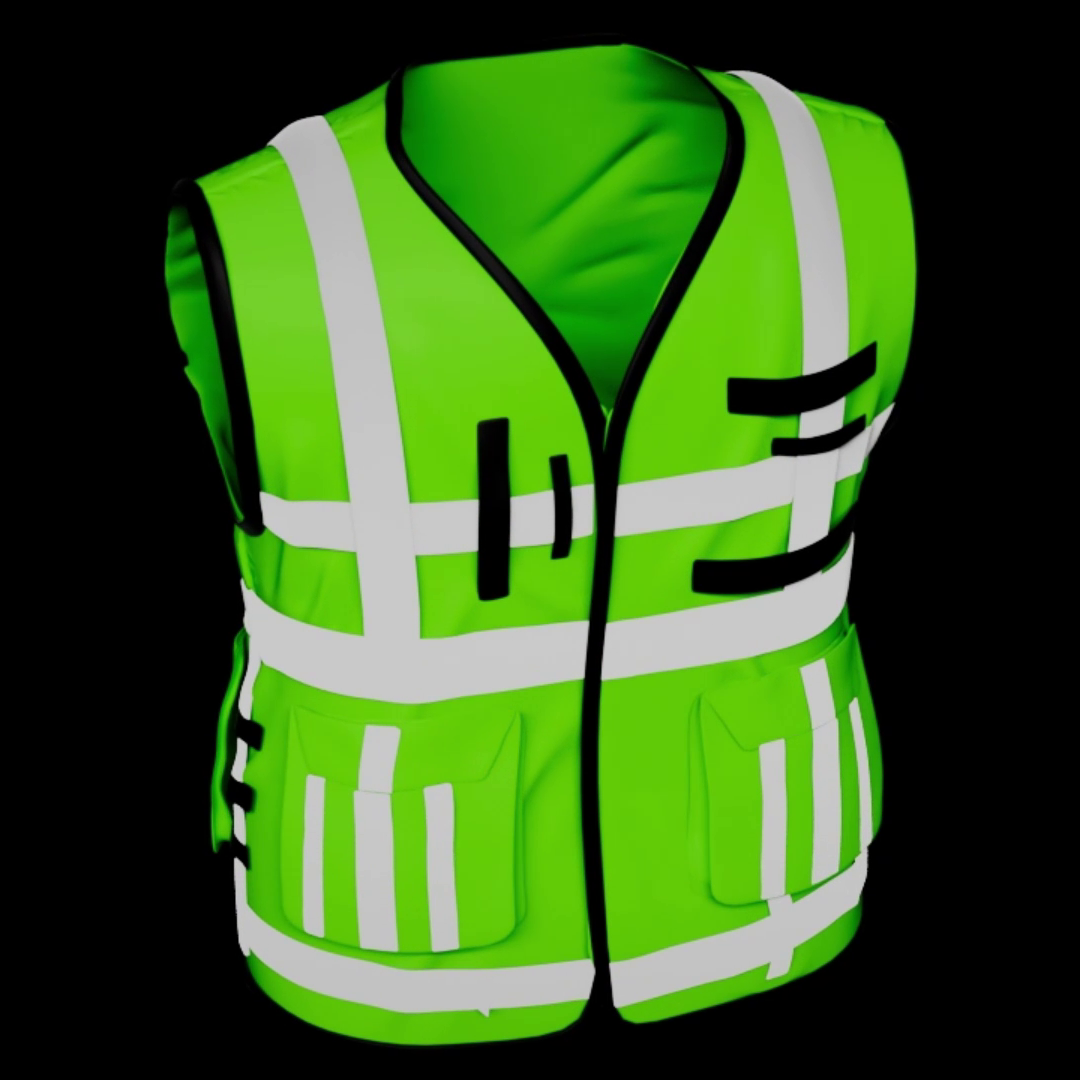}};
      \node[anchor=north west, text=white, font=\sffamily\bfseries\footnotesize, inner sep=3pt] at (img.north west) {MRM};
    \end{tikzpicture}%
    \begin{tikzpicture}[baseline=(img.south)]
      \node[anchor=south west, inner sep=0] (img) {\includegraphics[width=0.31\linewidth]{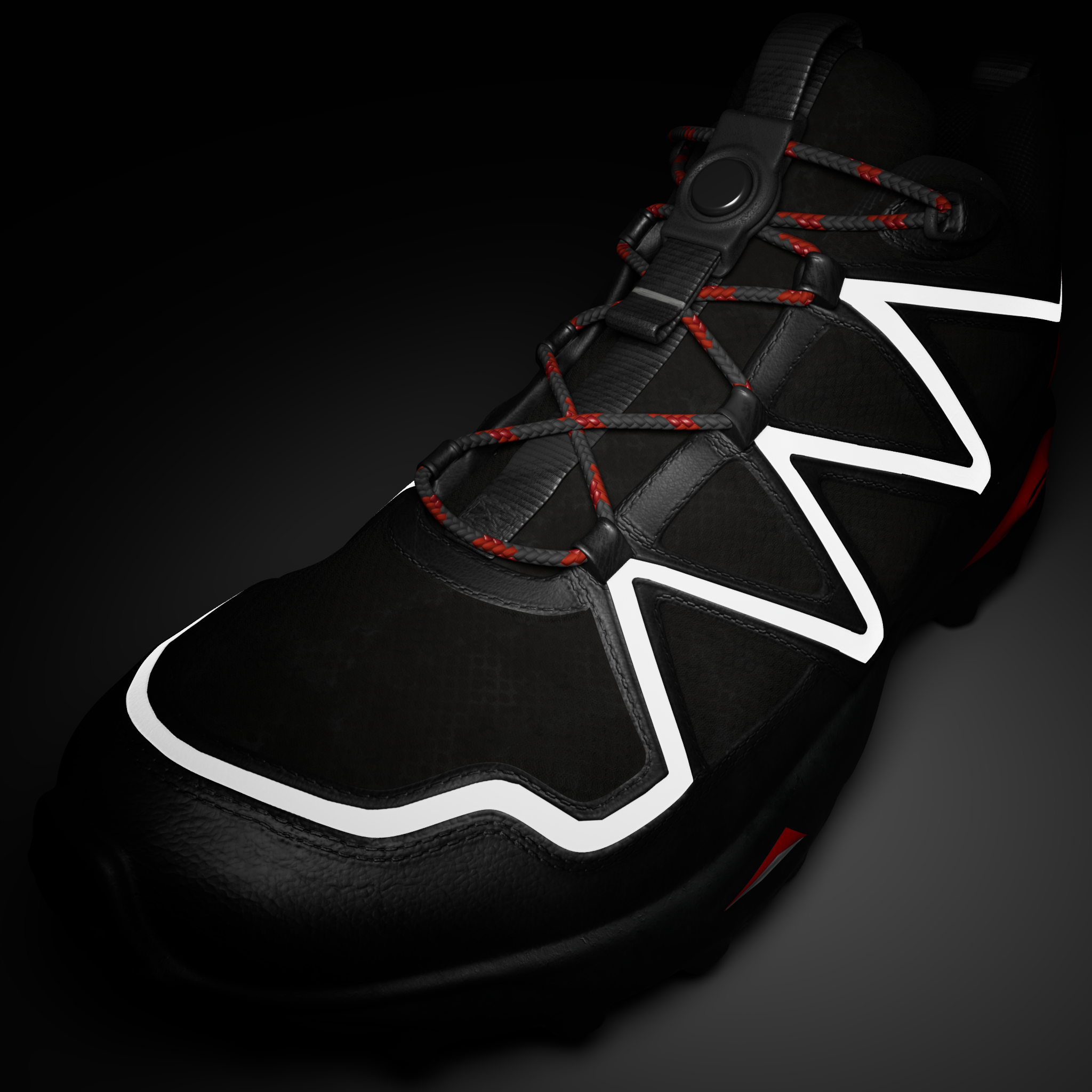}};
      \node[anchor=north west, text=white, font=\sffamily\bfseries\footnotesize, inner sep=3pt] at (img.north west) {MRM};
    \end{tikzpicture}
  \caption{Renderings of our MRM model (bottom row) obtained by an almost trivial modification of the equivalent roughness GGX microfacet model (top row). The lighting in all cases is aligned with the view direction, exhibiting a strong retroreflective peak in the MRM model.}
  \label{fig:teaser}
}

\maketitle

\begin{abstract}
  \small
  We present the Minimal Retroreflective Microfacet (MRM) model, which turns any existing microfacet BSDF into a physically plausible retroreflective one by a single substitution: replacing the view direction with its reflection about the surface normal before evaluating the standard model. Based on the previously published back-vector formulation, MRM requires only minimal code changes and has been adopted in the OpenPBR and MaterialX material standards. We prove reciprocity and energy conservation under the assumption of a reflection-symmetric normal distribution function (NDF), which holds for all commonly used distributions, and validate the model against measured retroreflective material data.
  \end{abstract}

\section{Introduction}

For safety applications such as road markings, signs, vehicles, and high-visibility clothing items, retroreflective materials are heavily used (see Figure~\ref{fig:teaser} for some example renderings). Materials are typically designed to be retroreflective via a manufactured substructure of elements (for example, sheets of micro-prisms or glass beads), which preferentially scatter light backwards \cite{Burgess2011}.
Realistic rendering of such materials is important for design and visualization purposes, as well as in a predictive rendering context, for example, for simulation of visibility in training of autonomous vehicles or for safety certification of road markings.

\paragraph{Previous work}
Modeling of retroreflectivity in the surface scattering function has received relatively little attention so far.
\citet{Neumann1999} presented a retroreflective modified Phong model; \citet{Edwards2006} proposed an empirical model that is not reciprocal; neither was compared to measurements.
\citet{Guo2017Retro} modeled retroreflection from prismatic sheets, but only for perfect retroreflection; their later extension~\cite{Guo2018Retro} adds a rough microfacet layer but lacks a closed-form expression.
\citet{Hericz2016} simulated glass bead retroreflection via ray tracing but also did not provide a closed-form BSDF.
The generalized cosine lobes of \citet{Lafortune1997} can represent retroreflectivity; they are reciprocal, and energy conservation can be enforced as the albedo is analytically computable. As a general-purpose fitting primitive, however, they lack intuitive controls such as a roughness parameter.
None of the aforementioned models can leverage existing microfacet infrastructure (importance sampling of visible normals, energy compensation, layering, pre-integration for real-time renderers) that MRM inherits directly.

\enlargethispage{0.1\baselineskip}

\paragraph{Our model}

For practical purposes in visual effects, we are interested in a model that is (a)~visually plausible and follows measured observations reasonably well, (b)~provably energy-conserving and reciprocal (the latter under the mild assumption of a reflection-symmetric NDF), (c)~easy to implement, (d)~compatible with existing microfacet infrastructure (such as image-based lighting pre-convolutions for real-time renderers), and (e)~efficient to run. We present here a model based on the previously published \emph{back-vector} formulation~\cite{BelcourBackvector}, which meets all of these requirements. It is so simple a modification to a regular microfacet model that we term it the Minimal Retroreflective Microfacet model, which achieves a plausible result.

In Section~\ref{sec:back_vector_microfacets}, we detail the derivation of MRM as a modification of the standard microfacet model~\cite{WalterMicrofacet,HeitzMicrofacet} using the back-vector as the microfacet normal selector. To check the plausibility of the model, we verify reciprocity symmetry in Section~\ref{sec:reciprocity} and energy conservation in Section~\ref{sec:energy_conservation}.
In Section~\ref{sec:implementation}, we sketch how the model can be implemented in practice, which is done trivially by reusing existing microfacet BSDF implementations. Finally in Section~\ref{sec:results}, we provide a comparison with measured data and some visual results.

\section{MRM as Back-Vector Modification of Microfacet Models}

\label{sec:back_vector_microfacets}

A microfacet BRDF \cite{WalterMicrofacet, HeitzMicrofacet} takes the form \citep[Equation (20)]{WalterMicrofacet}
\begin{equation}
  f_r(\vecv, \vecl) = \frac{D(\vechr) \, G_{2}(\vecv,\vecl,\vechr) \, F(\etav{\to}\etal, \vdot{\vecv}{\vechr})}{4 \avdot{\vecv}{\vecn} \avdot{\vecl}{\vecn}},
  \label{eq:microfacet_brdf}
\end{equation}
where $D$ is the distribution of micronormals (NDF), normalized to fulfill $\int_\Omega D(\vech) (\vdot{\vecn}{\vech}) \diff\vech = 1$, $G_{2}$ is the shadowing-masking function, and $F$ is the Fresnel term. The half-vector $\vechr = \vechr(\vecv, \vecl) = \frac{\vecv + \vecl}{\lVert \vecv + \vecl \rVert}$ is used as the \textit{selector} for microfacets reflecting light from incident direction $\vecl$ to outgoing view direction $\vecv$ (i.e.,\ only microfacets with normals aligned with the half-vector will reflect light).
Similarly, using the half-vector of refraction $\vecht = \vecht(\vecv, \vecl) = -{\left(\etav\vecv + \etal\vecl\right)}/{\lVert\etav\vecv + \etal\vecl\rVert}$ (where $\etav$, $\etal$ are the indices of refraction (IORs) of the media to which $\vecv$, $\vecl$ point) as the selector yields the corresponding BTDF:
\begin{equation}
  f_t(\vecv, \vecl) =
  \frac{\avdot{\vecv}{\vecht}\avdot{\vecl}{\vecht}\, \etav^2 \, D(\vecht) \, G_{2}(\vecv,\vecl,\vecht) \, \bigl(1 - F(\etav{\to}\etal, \vdot{\vecv}{\vecht}) \bigr)}
       {\avdot{\vecv}{\vecn} \avdot{\vecl}{\vecn} \left(\etav \vdot{\vecv}{\vecht} + \etal \vdot{\vecl}{\vecht} \right)^2}
  \label{eq:microfacet_btdf}.
\end{equation}
\citet[Sections~4.1 and 4.2]{WalterMicrofacet} detail the effect of using the half-vector as the microfacet selector and obtain Equations~\eqref{eq:microfacet_brdf} and \eqref{eq:microfacet_btdf}.
In this work, we show that using the \emph{back-vector}~\cite{BelcourBackvector} instead as a microfacet selector produces a highly plausible model of retroreflection.

To achieve retroreflection, in MRM we modify the microfacet model by changing the microfacet normal selector from the half-vector to the back-vector, defined as
\begin{equation}
  \vecbr(\vecv, \vecl) = \vechr(\vecvv, \vecl) = \frac{\vecvv + \vecl}{\lVert \vecvv + \vecl \rVert}, \quad \text{ with } \vecvv = \text{reflect}(\vecv, \vecn),
\end{equation}
to select the reflecting microfacet. 
For refraction, we can construct a similar refraction back-vector $\vecbt(\vecv, \vecl) = \vecht(\vecvv, \vecl)$.

\begin{figure}[!tb]
  \centering
    \includegraphics[width=0.48\textwidth]{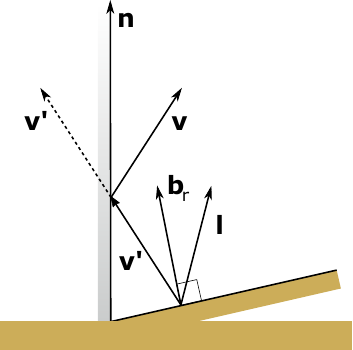} \hfil
    \includegraphics[width=0.48\textwidth]{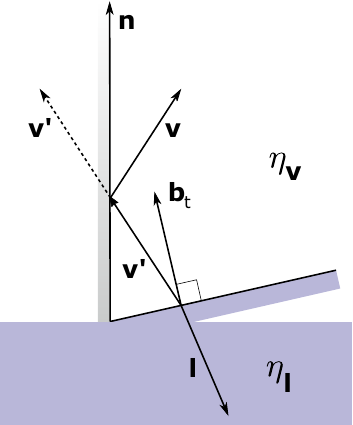}
  \caption{Geometry of the back-vector for reflection (left) and refraction (right) in MRM, illustrated in two dimensions. The view vector $\vecv$ is reflected in a ``virtual mirror'' to obtain $\vecvv$, which is then used to compute the half-vector with the light vector $\vecl$. The resulting back-vector ($\vecb_{r}$ for reflection and $\vecb_{t}$ for refraction) is used as the microfacet normal selector.}
  \label{fig:backvector}
\end{figure}


As illustrated in Figure \ref{fig:backvector}, the reflection $\vecvv$ of the view direction $\vecv$ on the surface can also be interpreted as a reflection of the view direction in a ``virtual mirror'' aligned with $\vecv$ and perpendicular to the surface macronormal $\vecn$, thus turning glossy forward-reflection into glossy retroreflection.  \citet{BelcourBackvector} showed that this reflection is related to the geometry of double scattering on the microsurface.

We emphasize that MRM is best understood as an \emph{empirical} model: unlike in the standard microfacet model, the back-vector $\vecbr$ does not correspond to a microfacet orientation from which a single specular reflection redirects light from $\vecl$ to $\vecv$. In practice, strong retroreflection arises from multiple-scattering substructures such as corner-reflector arrays or glass-bead layers, which are not directly modeled here. Rather, MRM repurposes the well-understood microfacet framework---its NDF lobe shape, masking-shadowing, and Jacobian---in the back-vector domain, producing a retroreflective lobe with the correct qualitative behavior.  The double-scattering interpretation of \citet{BelcourBackvector} provides physical intuition for why the microfacet components remain meaningful under this change of domain, and Section~\ref{sec:results} shows that this empirical approach provides a good match to measured retroreflective material data.

We now detail the mathematical consequences of using the back-vector as the microfacet selector.

\paragraph*{Jacobian of back-vector}
First we observe that the Jacobian of the change of variable from the back-vector to the light vector $\vecl$ for specular microfacets~\citep[Equations~(11) and~(14)]{WalterMicrofacet} has the same form as the half-vector one:
\begin{equation}
\left\lVert \dfrac{\partial \omega_{\vecb_{r}}}{\partial \omega_{\vecl}} \right\rVert = \left\lVert \dfrac{\partial \omega_{\vech_{r}(\vecvv, \vecl)}}{\partial \omega_{\vecl}} \right\rVert = \dfrac{1}{4 \left| \vdot{\vecl}{\vech_{r}(\vecvv, \vecl)} \right|} = \dfrac{1}{4 \left| \vdot{\vecl}{\vecbr} \right|}  .
\end{equation}
We also observe that this is true for refractive microfacets~\citep[Equation~(17)]{WalterMicrofacet}:
\begin{equation}
\left\lVert \dfrac{\partial \omega_{\vecb_{t}}}{\partial \omega_{\vecl}} \right\rVert = \left\lVert \dfrac{\partial \omega_{\vech_{t}(\vecvv, \vecl)}}{\partial \omega_{\vecl}} \right\rVert = \dfrac{  \etav^2 \avdot{\vecl}{\vech_{t}(\vecvv, \vecl)}}{ \bigl( \etav \vdot{\vecvv}{\vech_{t}(\vecvv, \vecl)} + \etal \vdot{\vecl}{\vech_{t}(\vecvv, \vecl)} \bigr)^2}  .
\end{equation}
The Jacobian has the same form as the half-vector case because we differentiate with respect to the light direction, which is not altered by the back-vector.

\paragraph*{Microfacet BSDF terms}
\hspace*{-1.3pt}The remaining ingredient is the Fresnel factor $\rho(\vecv,\vecm)$ \citep[Equation~(11)]{WalterMicrofacet}. Experimental evidence~\cite{BelcourBackvector} supports evaluating it at $\vecvv$ rather than $\vecv$, i.e.,\ substituting $\vecv \to \vecvv$ so that $\rho(\vecv,\vecm) = F(\etav{\to}\etal, \vdot{\vecvv}{\vecm})$.
For the reflection case~\citep[Equation~(15)]{WalterMicrofacet}, since $|\vdot{\vecl}{\vecbr}| = |\vdot{\vecvv}{\vecbr}|$---both equal the cosine of the half-angle between $\vecvv$ and $\vecl$---the Jacobian $1/(4|\vdot{\vecl}{\vecbr}|)$ simplifies identically to the standard derivation, giving
\begin{equation}
  f_r^m(\vecv, \vecl, \vecm) = F(\etav{\to}\etal, \vdot{\vecvv}{\vecm}) \dfrac{\delta(\vecb_r, \vecm) }{4 \left| \vdot{\vecvv}{\vecb_r} \right|^2}.
\end{equation}
For the transmissive case~\citep[Equation~(18)]{WalterMicrofacet}, the substitution $\vecht(\vecv,\vecl)$\linebreak $\to \vecbt = \vecht(\vecvv,\vecl)$ carries through directly in the refractive Jacobian, giving
\begin{equation}
  f_t^m(\vecv, \vecl, \vecm) = \left(1-F(\etav{\to}\etal, \vdot{\vecvv}{\vecm})\right) \dfrac{\delta(\vecb_t, \vecm) \etav^2 }{\left( \etav \vdot{\vecvv}{\vecb_t} + \etal \vdot{\vecl}{\vecb_t} \right)^2}.
\end{equation}
In both cases, the resulting macrosurface BSDF takes exactly the form of the standard microfacet model with $\vecv$ replaced by $\vecvv$: the NDF and Jacobian structure carry over unchanged, with $\vecvv$ appearing only in the Fresnel argument and the geometric denominator.
We note that $F$ here should be understood as an empirical attenuation rather than the physical Fresnel reflectance of a single smooth interface: since real retroreflective surfaces involve multiple internal scattering events (e.g.,\ corner-reflector arrays or glass-bead layers), the effective angular response may differ substantially from the standard IOR-based Fresnel factor, and flexible parameterizations such as $F_0$/$F_{82}$-tint (see Section~\ref{sec:results}) are well suited to fitting measured data.

\enlargethispage{1\baselineskip}

\paragraph*{Shadowing and masking}

The shadowing-masking function $G_{2}(\vecv, \vecl, \vech)$ accounts for the geometrical occlusion of the incident and outgoing rays by the local microsurface geometry. It combines the visibility masking function $G_1(\vecv, \vech)$ for the view direction and the visibility shadowing function $G_1(\vecl, \vech)$ for the light direction. Popular choices for $G_1$
are the Smith masking function\footnote{Where $\chi^+$ is the Heaviside step function, and $\cldot{\vecv}{\vech}$ denotes $\vdot{\vecv}{\vech}$ clamped to be nonnegative.}
\begin{equation} \label{eq:smith_masking}
  G_{1,\text{Smith}}(\vecv, \vech) = \frac{\chi^+(\vdot{\vecv}{\vech})  \avdot{\vecv}{\vecn}}{\int \cldot{\vecv}{\vech} D(\vech) \diff\vech},
\end{equation}
where analytic expressions are available for Beckmann and GGX distributions, and the V-cavity masking function \cite{HeitzMicrofacet}
\begin{equation} \label{eq:vc_masking}
  G_{1,\text{VC}}(\vecv, \vech) =
  \min\left\lbrace\frac{2\avdot{\vecv}{\vecn}\avdot{\vecn}{\vech}}{\avdot{\vecv}{\vech}}, 1\right\rbrace,
\end{equation}
which is generally applicable for distributions with symmetry around the normal.

The shadowing-masking function $G_{2}(\vecv, \vecl, \vech)$ combines the shadowing and masking functions in a reciprocal manner, for example in the separable form commonly used in microfacet models $G_{2}(\vecv, \vecl, \vech) = G_1(\vecv, \vech) \, G_1(\vecl, \vech)$; more generally, $G_2$ can be any function of $G_1(\vecv, \vech)$ and $G_1(\vecl, \vech)$ satisfying reciprocity \citep{HeitzMicrofacet}.

We assume Smith or V-cavity masking is applied to the mirrored microfacet distribution, and thus replace  $G_{2}(\vecv, \vecl, \vech)$  with $G_{2}(\vecvv, \vecl, \vech)$ in the retroreflective microfacet model. With Smith masking under the assumption of a reflection-symmetric NDF, this is actually identical to a regular microfacet BSDF, as $\vecvv$ and $\vecv$ share the same geometric configuration (see Section~\ref{sec:reciprocity}). For V-cavity masking, taking $\vecvv$ as the view direction simply corresponds to selecting $\vecb$ as the microfacet normal.

\paragraph*{Final form}
It follows from the above that MRM reduces to simply replacing $\vecv$ by $\vecvv$ in the standard microfacet BRDF $f_r(\vecv,\vecl)$ and BTDF $f_t(\vecv,\vecl)$:
\vspace*{-3pt}\begin{equation} \label{eq:retro_microfacet_model}
  {\setlength{\fboxsep}{5pt}\setlength{\fboxrule}{1pt}%
  \fcolorbox{red}{white}{$\displaystyle f_\text{MRM}(\vecv, \vecl) =
  \begin{cases}
    f_{r, \text{MRM}}(\vecv, \vecl) = f_r(\vecvv, \vecl)  & \text{(reflection, \;} (\vdot{\vecv}{\vecn})(\vdot{\vecl}{\vecn}) \geq 0\text{)}, \\
    f_{t, \text{MRM}}(\vecv, \vecl) = f_t(\vecvv, \vecl)  & \text{(transmission, otherwise)}.
  \end{cases}$}}
\end{equation}
\vspace*{-9pt}

\begin{figure}[tb]
  \centering
  \includegraphics[width=0.99\linewidth]{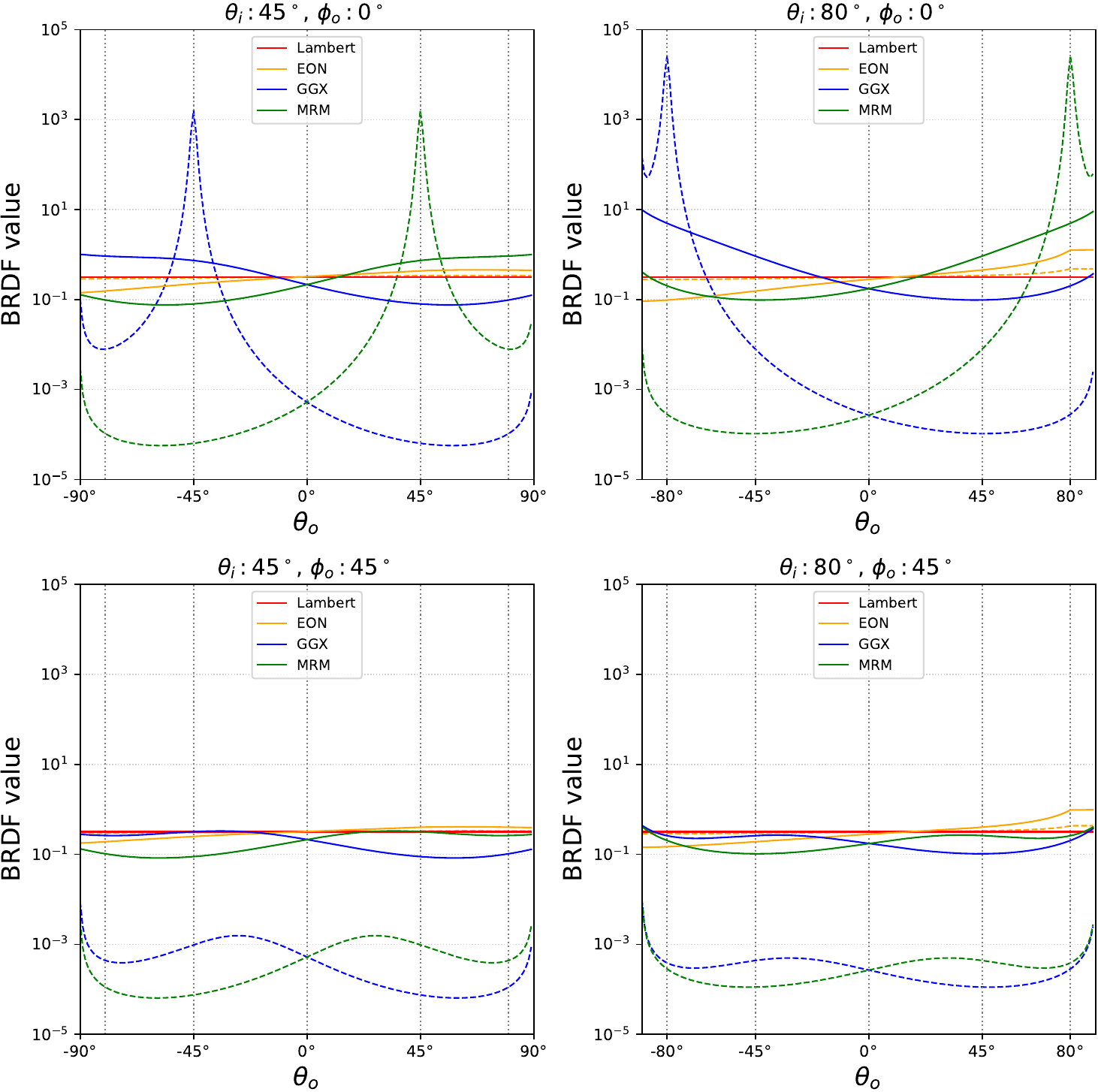}
  \caption{Comparison of the retroreflective MRM BRDF (green) to the regular GGX microfacet BRDF (blue) and rough-diffuse EON model (orange), as a function of view direction $\theta_o$ for fixed light directions $\theta_i \in \{45^\circ, 80^\circ\}$. The dashed lines are the roughness $0.1$ case, and the solid lines roughness $0.666$. (The horizontal red line at $1/\pi$ is the Lambert BRDF.)}
  \label{fig:brdf_plots}
  \vspace*{-6pt}
\end{figure}

\paragraph*{BRDF comparison}
Figure~\ref{fig:brdf_plots} plots the MRM (green), GGX (blue), and rough-diffuse EON~\cite{Portsmouth2025EON} (orange) models, with Fresnel set to 1 for both MRM and GGX.

MRM has the same lobe shape as GGX, but with the peak flipped to the exact retroreflection direction ($\theta_o = \theta_i$, $\phi_o = \phi_i$). Low roughness ($0.1$) produces a very tight, bright peak (note the logarithmic $y$-axis), while high roughness ($0.666$) gives a broad, dimmer one.

The $\phi_o = 45^\circ$ case shows behavior when the light and view vectors are not coplanar: both GGX and MRM fall well below Lambert because at those directions almost no energy is scattered---it is concentrated instead in the specular or retroreflective peak, respectively.

\enlargethispage{.5\baselineskip}

\section{Plausibility of MRM}

We show here that the retroreflective microfacet model of Equation~\eqref{eq:retro_microfacet_model} is physically plausible, i.e.,\ that it fulfills reciprocity symmetry and energy conservation requirements.


\subsection{Reciprocity}

\label{sec:reciprocity}

We first verify that the MRM model fulfills reciprocity symmetry, i.e.,\
\begin{equation}
  f_{\text{MRM}}(\vecv, \vecl) = f_{\text{MRM}}(\vecl, \vecv)  , \quad f_{t,\text{MRM}}(\vecv, \vecl) = \frac{\etav^2}{\etal^2}f_{t,\text{MRM}}(\vecl, \vecv)  .
\end{equation}
As outlined by \citet{Arvo95}, we can express the reflection around the normal using a matrix transform $\Rn$:
\begin{equation}
  \text{reflect}(\vecx, \vecn) = \Rn \vecx = \left(2 \vecn\vecn^T - \mathbf{I} \right) \vecx,
\end{equation}
where $\mathbf{I}$ is the identity matrix.
Note that $\Rn$ is the Householder matrix multiplied by $-1$, which means some of its properties are applicable to $\Rn$.
The Householder matrix is orthogonal, which implies that dot products do not change, so we have $\vdot{\Rn \vecx}{\Rn \vecy} = \vdot{\vecx}{\vecy}$ and, in turn, the norm is preserved ($\lVert \Rn \vecx \rVert = \lVert \vecx \rVert$).
Further, like the Householder matrix, $\Rn$ is involutory, i.e.,\ it is its own inverse ($\Rn \Rn \vecx = \vecx$).
\enlargethispage{-0.5\baselineskip}
Using those properties of $\Rn$, we now show that the back-vector gets reflected under the swap $(\vecv, \vecl) \leftrightarrow (\vecl, \vecv)$:\enlargethispage{.5\baselineskip}
\begin{eqnarray}\label{eq:backvector_swap}
  \begin{aligned}
    \Rn \vecbr(\vecv, \vecl)
    &= \Rn \vechr(\vecvv, \vecl)
    = \Rn \frac{\vecvv + \vecl}{\lVert \vecvv + \vecl \rVert}\\
    &= \frac{\Rn \vecvv + \Rn \vecl}{\lVert \cdot \rVert}
    = \frac{\vecv + \vecll}{\lVert \cdot \rVert}
    = \vechr(\vecll, \vecv)
    = \vecbr(\vecl, \vecv),
  \end{aligned}
\end{eqnarray}
where we omitted spelling out the norm in the denominator (as it continues matching the norm of the numerator, which is unchanged due to norm-preservation) and adopted the notation $\vecll$ for the reflected light direction.
Analogously, the same can be shown for $\vecbt$, and in the following we use $\vecb$ as all equations apply for both $\vecbr$ and $\vecbt$.

\paragraph*{NDF term}

We will need one technical assumption on the NDF $D$: we require that $D$ is \emph{reflection-symmetric} around the normal.
Note that this is also the requirement for the applicability of V-cavity masking and that it holds for all commonly used distributions including isotropic and anisotropic GGX, Beckmann, and Phong.
Under that requirement for $D$ and using Equation~\eqref{eq:backvector_swap} we immediately conclude $D\left(\vecb(\vecv, \vecl)\right) = D\left(\Rn\vecb(\vecv, \vecl)\right) = D\left(\vecb(\vecl, \vecv)\right)$.

\paragraph*{Shadowing and masking}

For V-cavity masking, using Equation~\eqref{eq:backvector_swap}, the orthogonality of $\Rn$, and observing that $\Rn\vecn = \vecn$, we get
\begin{eqnarray}
  \begin{aligned}
    G_{1,\text{VC}}(\vecvv, \vecb(\vecv, \vecl))
    &= \min\left\lbrace\frac{2\avdot{\vecvv}{\vecn}\avdot{\vecn}{\vecb(\vecv, \vecl)}}{\avdot{\vecvv}{\vecb(\vecv, \vecl)}}, 1\right\rbrace\\
    &= \min\left\lbrace\frac{2\avdot{\Rn\vecvv}{\Rn\vecn}\avdot{\Rn\vecn}{\Rn\vecb(\vecv, \vecl)}}{\avdot{\Rn\vecvv}{\Rn\vecb(\vecv, \vecl)}}, 1\right\rbrace\\
    &= \min\left\lbrace\frac{2\avdot{\vecv}{\vecn}\avdot{\vecn}{\vecb(\vecl, \vecv)}}{\avdot{\vecv}{\vecb(\vecl, \vecv)}}, 1\right\rbrace
    = G_{1,\text{VC}}(\vecv, \vecb(\vecl, \vecv)).
  \end{aligned}
\end{eqnarray}
Similarly, using the same properties, for Smith masking we obtain
\begin{eqnarray}
  \begin{aligned}
    G_{1,\text{Smith}}\left(\vecvv, \vecb(\vecv, \vecl)\right) &=
    \frac{\chi^+\left(\vdot{\vecvv}{\vecb(\vecv, \vecl)}\right) \avdot{\vecvv}{\vecn}}{\int \cldot{\vecvv}{\vecx} D(\vecx) \diff\vecx}\\
    &= \frac{\chi^+\left(\vdot{\Rn\vecvv}{\Rn\vecb(\vecv, \vecl)}\right) \avdot{\Rn\vecvv}{\Rn\vecn}}{\int \cldot{\Rn\vecvv}{\Rn\vecx} D(\Rn\vecx) \diff\vecx}\\
    &= \frac{\chi^+\left(\vdot{\vecv}{\vecb(\vecl, \vecv)}\right) \avdot{\vecv}{\vecn}}{\int \cldot{\vecv}{\Rn\vecx} D(\Rn\vecx) \diff\vecx}\\
    &= \frac{\chi^+\left(\vdot{\vecv}{\vecb(\vecl, \vecv)}\right) \avdot{\vecv}{\vecn}}{\int \cldot{\vecv}{\vecx} D(\vecx) \diff\vecx} = G_{1,\text{Smith}}\left(\vecv, \vecb(\vecl, \vecv)\right),
  \end{aligned}
\end{eqnarray}
where in the last step we simply changed the integration domain to the $\Rn$-transformed hemisphere of directions.

In analogy to the above, we obtain $G_1\left(\vecl, \vecb(\vecv, \vecl)\right) = G_1\left(\vecll, \vecb(\vecl, \vecv)\right)$ for both masking variants. For a separable $G_2 = G_1 \cdot G_1$ (or any $G_2$ symmetric in its first two arguments), combining both $G_1$ results gives $G_2\!\left(\vecvv, \vecl, \vecb(\vecv, \vecl)\right) = G_2\!\left(\vecv, \vecll, \vecb(\vecl, \vecv)\right) = G_2\!\left(\vecll, \vecv, \vecb(\vecl, \vecv)\right)$, which completes the proof of reciprocity of the shadowing-masking term.

\paragraph*{Fresnel term}


For \emph{reflection}, $f_r(\vecvv,\vecl)$ evaluates $F(\etav{\to}\etal, \vdot{\vecvv}{\vecbr(\vecv,\vecl)})$ and the swap\-ped path $F(\etav{\to}\etal, \vdot{\vecll}{\vecbr(\vecl,\vecv)})$; since $\vecbr(\vecv,\vecl)\propto\vecvv+\vecl$, the forward cosine is \mbox{$(1+\vdot{\vecvv}{\vecl})/\|\vecvv+\vecl\|$}, and $\vecbr(\vecl,\vecv)\propto\vecll+\vecv$ gives the same expression by $\Rn$-orthog\-o\-nal\-ity ($\vdot{\vecvv}{\vecl}=\vdot{\vecll}{\vecv}$).
For \emph{refraction}, denote by $T_{a\to b}$ the Fresnel transmittance of a ray from $a$ to $b$ through the microfacet. Ray $\vecvv$ refracts through $\vecbt(\vecv,\vecl)$ to $\vecl$ (transmittance $T_{\vecvv\to \vecl}$); the swapped path has $\vecll$ refracting through $\vecbt(\vecl,\vecv)$ to $\vecv$ (transmittance $T_{\vecll\to \vecv}$). By reciprocity of light transport, $T_{\vecvv\to \vecl} = T_{\vecl\to \vecvv}$. Via Equation~\eqref{eq:backvector_swap}, $\vdot{\vecl}{\vecbt(\vecv,\vecl)} = \vdot{\vecll}{\vecbt(\vecl,\vecv)}$, so $\vecl$ and $\vecll$ hit their microfacets at the same angle across the same $\etal{\to}\etav$ boundary, hence $T_{\vecl\to \vecvv} = T_{\vecll\to \vecv}$. Combining, $T_{\vecvv\to \vecl} = T_{\vecll\to \vecv}$.

\paragraph*{Summary}

In summary, we can conclude the reciprocity symmetry of the BRDF:
\begin{equation}
  f_{r,\text{MRM}}(\vecv, \vecl) = f_r(\vecvv, \vecl) =f_r(\vecll, \vecv) =  f_{r,\text{MRM}}(\vecl, \vecv).
\end{equation}
Analogously, we obtain the reciprocity symmetry requirement of the BTDF:
\begin{equation}
    f_{t,\text{MRM}}(\vecv, \vecl) =  f_t(\vecvv, \vecl) = \frac{\etav^2}{\etal^2}f_t(\vecll, \vecv) = \frac{\etav^2}{\etal^2}f_{t,\text{MRM}}(\vecl, \vecv).
\end{equation}


\subsection{Energy Conservation}

\label{sec:energy_conservation}

We now verify that the retroreflective microfacet model fulfills energy conservation, i.e.,\ that the directional albedos fulfill
$\rho_{r, \text{MRM}}(\vecv) + \rho_{t, \text{MRM}}(\vecv) \leq 1$ for all $\vecv$.
For the directional albedos we simply have (where $\Omega_{\vecx}$ denotes the hemisphere containing $\vecx$)
\begin{align}
  \rho_{r, \text{MRM}}(\vecv)
    &= \int_{\Omega_{\vecv}} f_{r, \text{MRM}}(\vecv, \vecl) \avdot{\vecl}{\vecn} \diff\vecl
     = \int_{\Omega_{\vecvv}} f_r(\vecvv, \vecl) \avdot{\vecl}{\vecn} \diff\vecl
     = \rho_r(\vecvv), \nonumber\\[4pt]
  \rho_{t, \text{MRM}}(\vecv)
    &= \int_{\Omega_{-\vecv}} f_{t, \text{MRM}}(\vecv, \vecl) \avdot{\vecl}{\vecn} \diff\vecl
     = \int_{\Omega_{-\vecvv}} f_t(\vecvv, \vecl) \avdot{\vecl}{\vecn} \diff\vecl
     = \rho_t(\vecvv).
\end{align}
That is, the directional albedo of the MRM is equivalent to the directional albedo of the regular microfacet BSDF, but evaluated at the reflected direction.
As the original microfacet BSDF fulfills energy conservation, the retroreflective modification does, too.

We further note that under our requirement of a reflection-symmetric microfacet normal distribution, the geometric
configurations for $\vecv$ and $\vecvv$ are identical.
Hence we can conclude $\rho_r(\vecv) = \rho_r(\vecvv)$ and $\rho_t(\vecv) = \rho_t(\vecvv)$, which in turn implies that the albedos of forward- and retroreflection are the same, i.e.,\ $\rho_{r, \text{MRM}}(\vecv) = \rho_r(\vecv)$ and $\rho_{t, \text{MRM}}(\vecv) = \rho_t(\vecv)$.

\section{Implementation Notes}
\label{sec:implementation}

Given an implementation of a regular microfacet BSDF, extending it to retroreflection is extremely straightforward (see the pseudocode provided in Listing~\ref{lst:implementation}):

\begin{itemize}
\item
  Evaluation merely needs to replace $\vecv$ with $\vecvv$ upfront.
\item
  Similarly, importance sampling of $\vecl$ given $\vecv$ can be realized by replacing $\vecv$ with $\vecvv$ upfront and then importance sampling the regular microfacet BSDF.
  This may include low variance sampling using the domain of visible microfacets \cite{HeitzIS}.
  The sampling PDF is likewise the regular microfacet PDF evaluated with $\vecvv$ in place of $\vecv$.
\item
  As the albedos of standard BSDF and the retroreflective BSDF are identical, compensating for the energy loss of microfacet single-scattering via a diffuse term \cite{KelemenBRDF, Kulla17} or normalization \cite{TurquinMultipleScattering} can thus be realized using the same data tables. The exact energy compensation of \citet{Heitz2016} can also be applied unchanged.
\item An implementation may optionally wish to provide a blend between regular and retroreflective microfacet behavior to model materials with partial retroreflective properties (for example this is required in OpenPBR \cite{OpenPBR2024}). This can be achieved simply by blending $f_{r, \text{MRM}}$ and $f_r$, or alternatively via a stochastic selection (e.g.,\ one-sample MIS \cite{Veach1998}) between the original $\vecv$ and reflected $\vecvv$ view directions during sampling.
\end{itemize}

\begin{lstlisting}[
 %   style=snippet,
    language=GLSL,
    caption={Pseudocode for the MRM model: trivial wrappers around an existing microfacet BSDF implementation, substituting $\vecvv$ for $\vecv$.},
    label=lst:implementation,
%    nolol=true,
%    frame=trBL,
    float=b,
    abovecaptionskip=5pt,
%    aboveskip=4pt,
%    belowskip=4pt
    ]
Sample_result sample_bsdf_retro(vec3 V, vec3 N, Bsdf_params params)
{
    vec3 Vp = -V + 2.0*dot(V, N)*N;  // Reflected view direction
    return sample_bsdf(Vp, N, params);  // {L, f, pdf}
}

Eval_result eval_bsdf_retro(vec3 V, vec3 L, vec3 N, Bsdf_params params)
{
    vec3 Vp = -V + 2.0*dot(V, N)*N;  // Reflected view direction
    return eval_bsdf(Vp, L, N, params);
}

float pdf_bsdf_retro(vec3 V, vec3 L, vec3 N, Bsdf_params params)
{
    vec3 Vp = -V + 2.0*dot(V, N)*N;  // Reflected view direction
    return pdf_bsdf(Vp, L, N, params);
}
\end{lstlisting}

\section{Results and Discussion}

\label{sec:results}

\begin{figure}[tb]
  \centering
  \begin{tikzpicture}[baseline=(img.south)]
    \node[anchor=south west, inner sep=0] (img) {\includegraphics[height=2.8cm,frame]{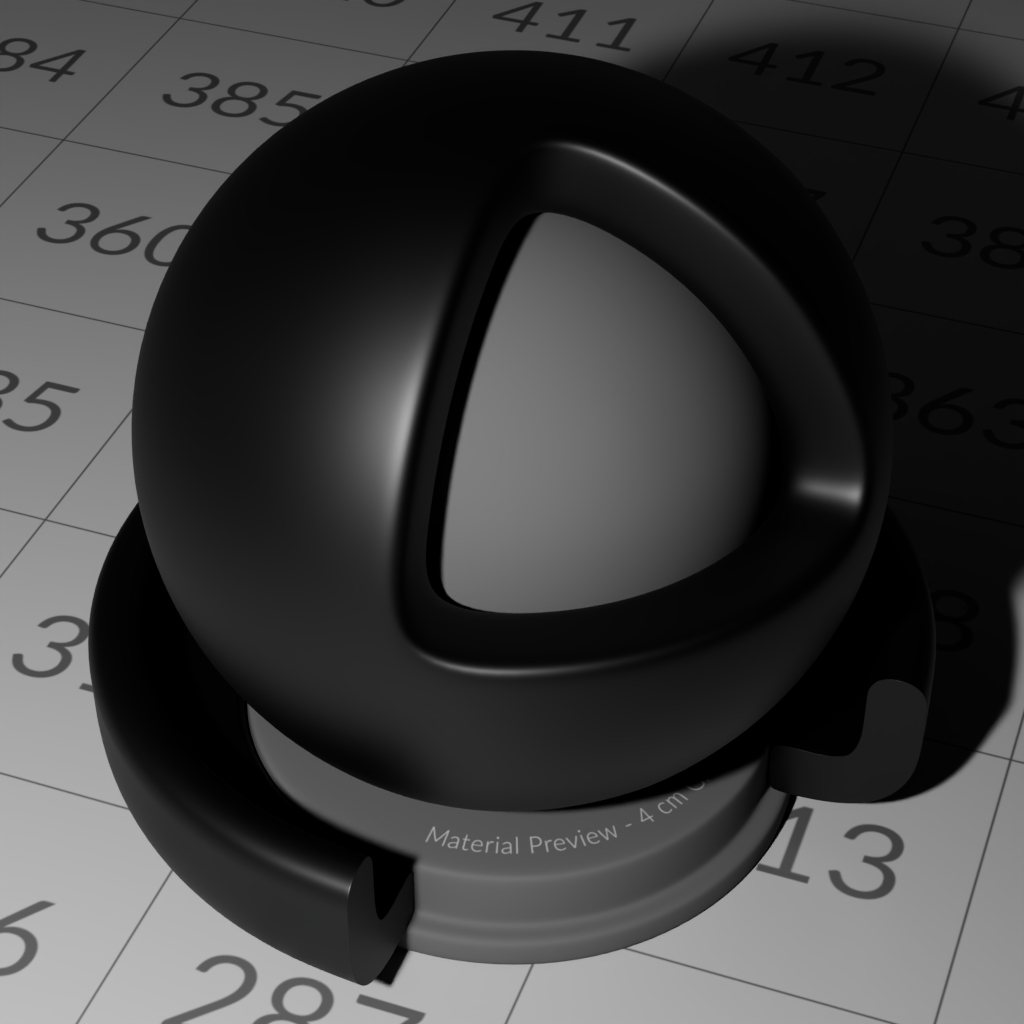}};
    \node[anchor=north west, text=white, font=\sffamily\bfseries\footnotesize, inner sep=3pt] at (img.north west) {GGX};
  \end{tikzpicture}%
  \hspace{0.2cm}%
  \begin{tikzpicture}[baseline=(img.south)]
    \node[anchor=south west, inner sep=0] (img) {\includegraphics[height=2.8cm,frame]{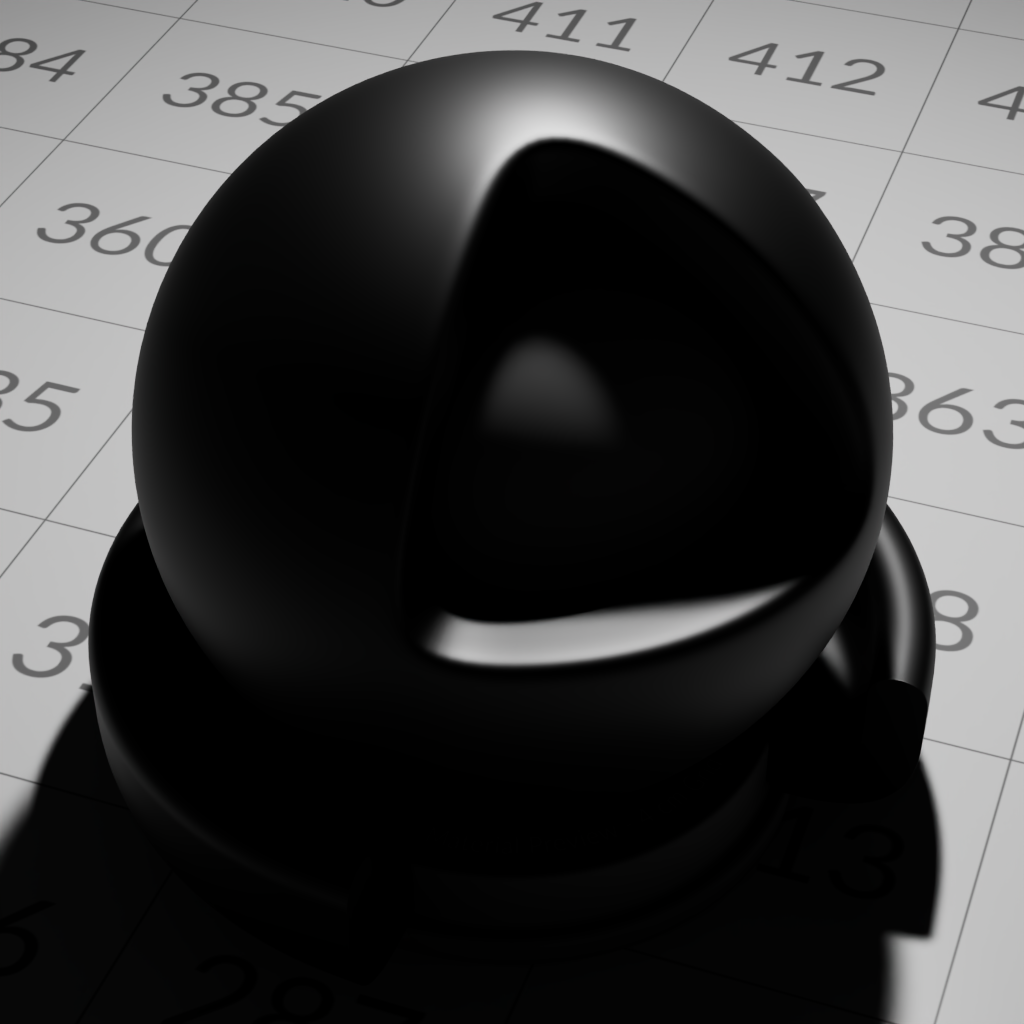}};
    \node[anchor=north west, text=white, font=\sffamily\bfseries\footnotesize, inner sep=3pt] at (img.north west) {GGX};
  \end{tikzpicture}%
  \hspace{0.2cm}%
  \begin{tikzpicture}[baseline=(img.south)]
    \node[anchor=south west, inner sep=0] (img) {\includegraphics[height=2.8cm,frame]{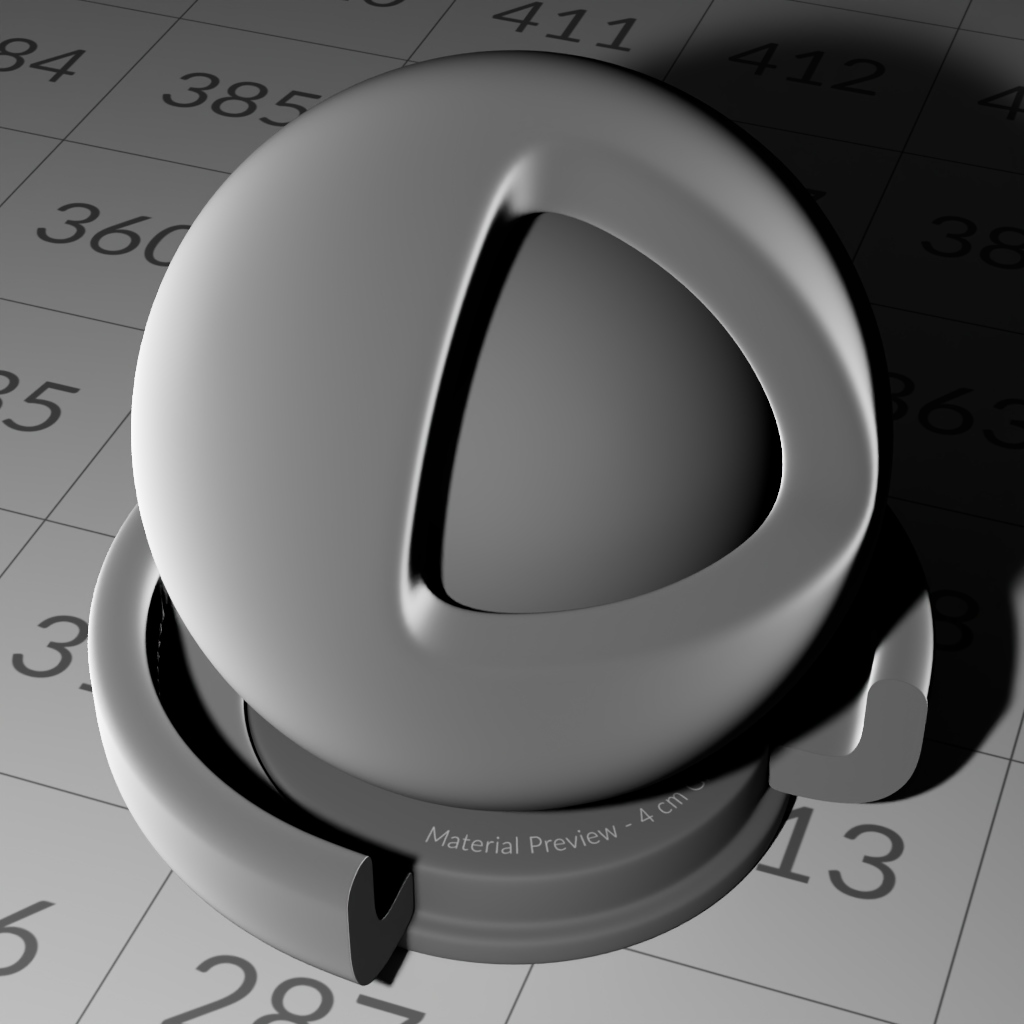}};
    \node[anchor=north west, text=white, font=\sffamily\bfseries\footnotesize, inner sep=3pt] at (img.north west) {MRM};
  \end{tikzpicture}%
  \hspace{0.2cm}%
  \begin{tikzpicture}[baseline=(img.south)]
    \node[anchor=south west, inner sep=0] (img) {\includegraphics[height=2.8cm,frame]{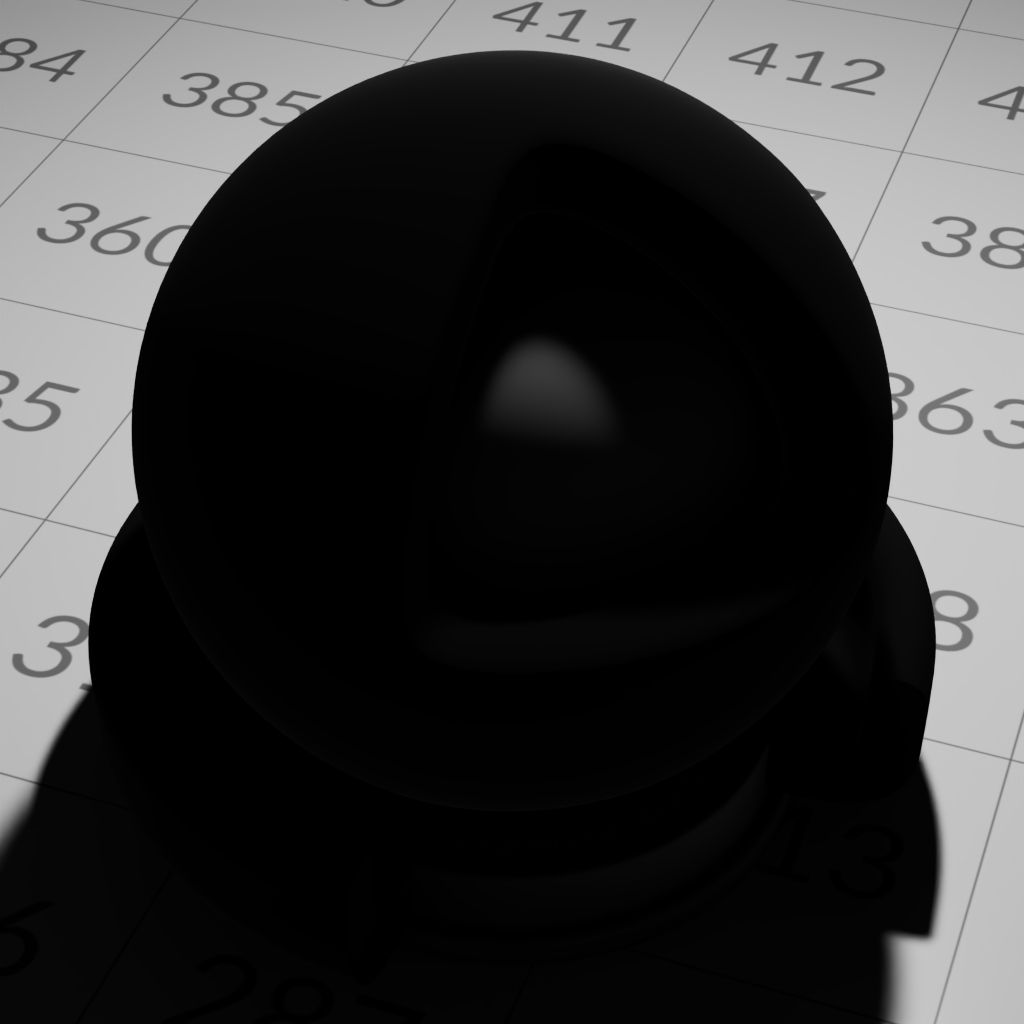}};
    \node[anchor=north west, text=white, font=\sffamily\bfseries\footnotesize, inner sep=3pt] at (img.north west) {MRM};
  \end{tikzpicture}
  \caption{Specular shaderball (non-metallic, roughness $0.5$) with GGX (left two) and MRM (right two), each shown with $\vecv \approx \vecl$ and $\vecv \not\approx \vecl$. When $\vecv \approx \vecl$, the MRM highlight spans the entire surface because $\vecbr \approx \vecn$ everywhere; moving the light away causes it to vanish, whereas the GGX highlight merely shifts.}
  \label{fig:highlight_extent}
\end{figure}

\paragraph*{Highlight appearance on curved surfaces}
Although the MRM BRDF has the same lobe shape as GGX at any single surface point, the apparent highlight on \emph{curved} surfaces is significantly broader under retroreflection (Figure~\ref{fig:highlight_extent}).
When $\vecv = \vecl$, the back-vector $\vecbr = \vecn$ exactly at every surface point, so $D(\vecbr)$ stays at its peak across the entire visible surface; the half-vector, by contrast, is approximately constant for distant illumination, so $D(\vechr)$ peaks only where $\vecn \approx \vechr$.
This is a fundamental property of retroreflective materials in general, not specific to MRM.

Energy is conserved---MRM merely redirects each point's specular lobe toward the camera (Section~\ref{sec:energy_conservation}).
From an artist's perspective, roughness controls how closely the light source must be aligned with the camera to produce the retroreflective effect, \emph{not} the apparent size of the highlight.

\paragraph*{Comparison to measured data}

\begin{figure}[!b]
    \includegraphics[width=\textwidth]{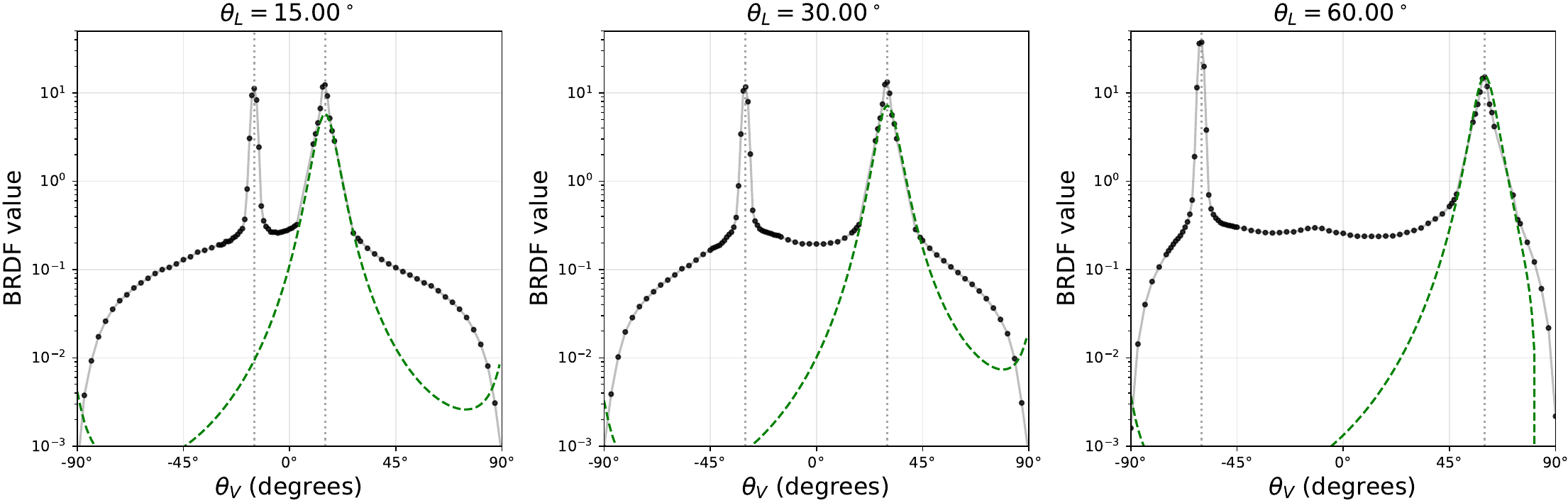}
  \caption{Measured BRDF of retroreflective tape (black dots) as a function of view angle $\theta_V$, at three different light source angles $\theta_L \in \{15^\circ, 30^\circ, 60^\circ\}$, and the MRM model fit (green line).}
  \label{fig:data_fit}
\end{figure}

Figure~\ref{fig:data_fit} shows a fit of the retroreflective MRM model (green dashed line) to the measured BRDF of retroreflective tape (black dots) from \cite{BelcourBackvector}. This is the ``yellow tape'' sample from their dataset, illuminated at three different light source angles $\theta_L \in \{15^\circ, 30^\circ, 60^\circ\}$ (in the plane of incidence).  The MRM model provides a reasonably good match to the retroreflective peak using a roughness of $r = 0.23$, together with ``F82-tint'' Fresnel \cite{OpenPBR2024} with $F_0=0.19$ and $F_{82} = 0$.

\enlargethispage{0.5\baselineskip}

\paragraph*{Incorporation in a standard microfacet model}

\begin{figure}
  \centering
  \begin{tikzpicture}[font=\scriptsize ]
    \node[inner sep=1pt] (GGX0) {\includegraphics[width=0.3\linewidth,frame]{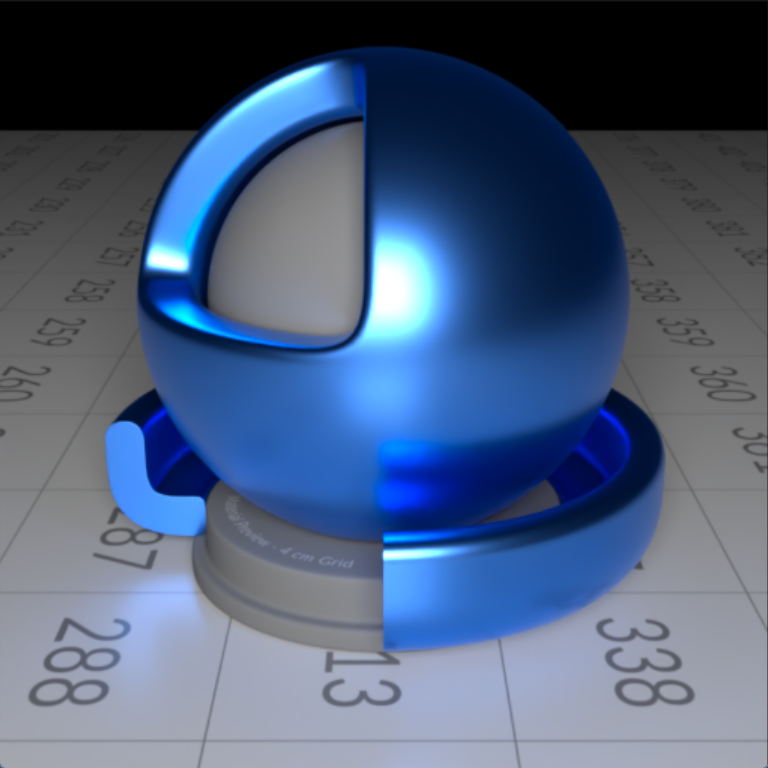}};
    \node[inner sep=1pt,right = 0cm of GGX0] (GGX45) {\includegraphics[width=0.3\linewidth,frame]{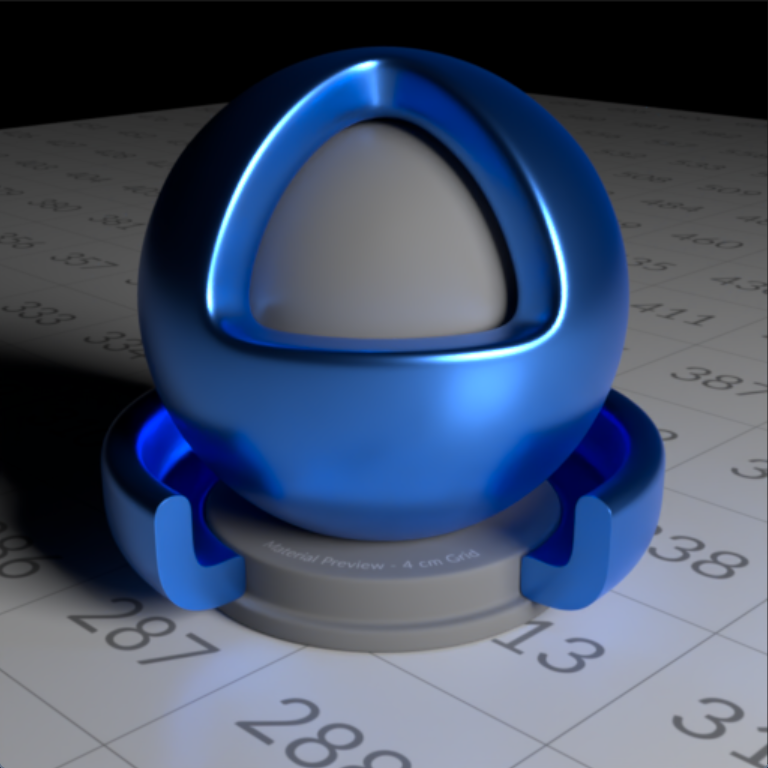}};
    \node[inner sep=1pt,right = 0cm of GGX45] (GGX90) {\includegraphics[width=0.3\linewidth,frame]{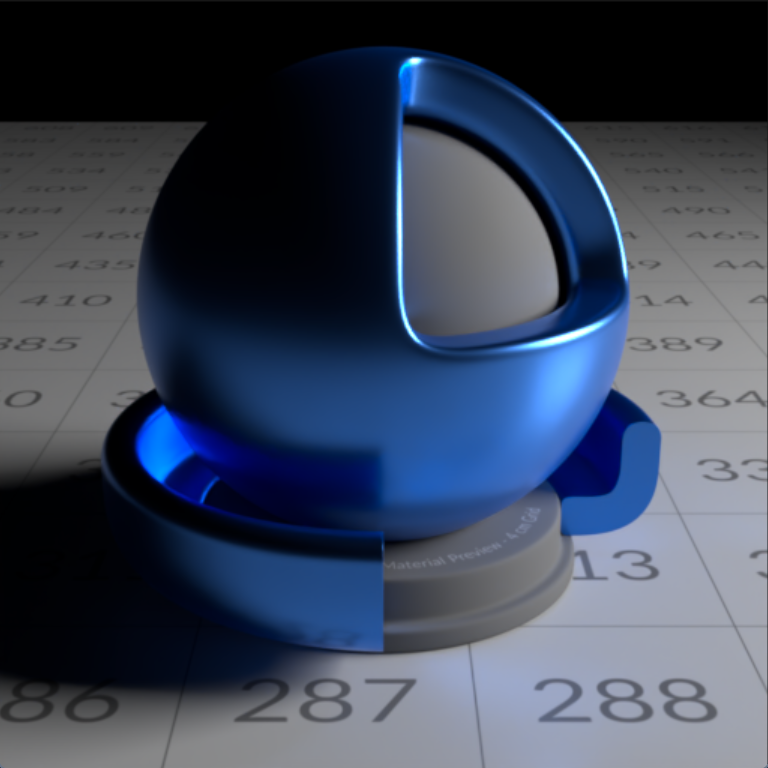}};
    \node[inner sep=1pt,below = 0cm of GGX0] (MRM0) {\includegraphics[width=0.3\linewidth,frame]{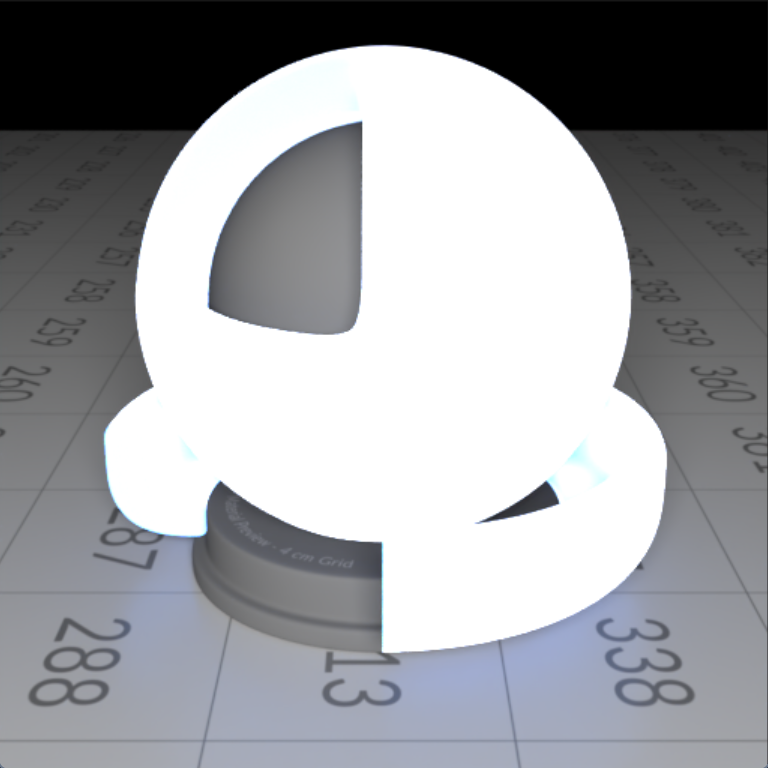}};
    \node[inner sep=1pt,right = 0cm of MRM0] (MRM45) {\includegraphics[width=0.3\linewidth,frame]{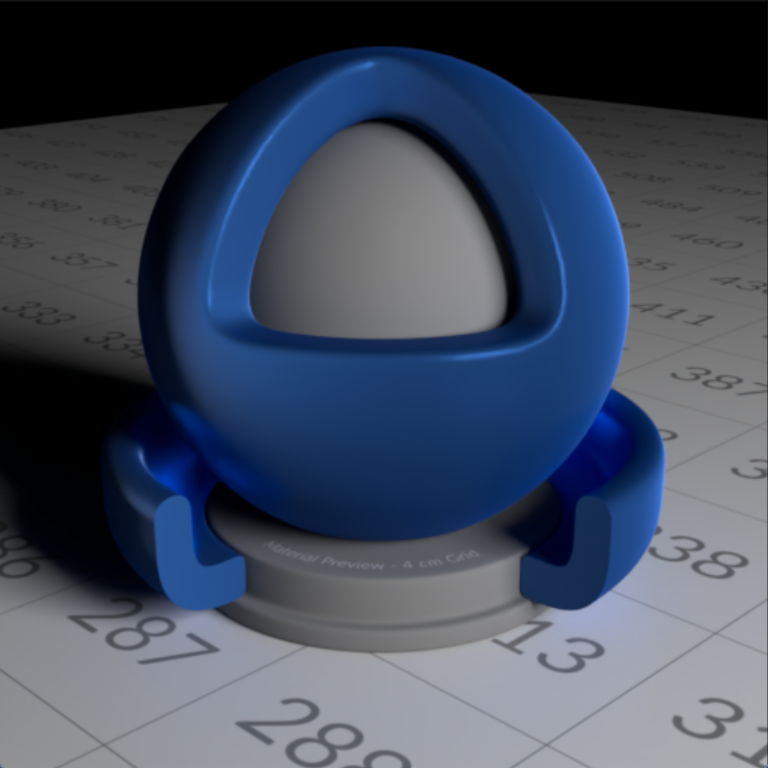}};
    \node[inner sep=1pt,right = 0cm of MRM45] (MRM90) {\includegraphics[width=0.3\linewidth,frame]{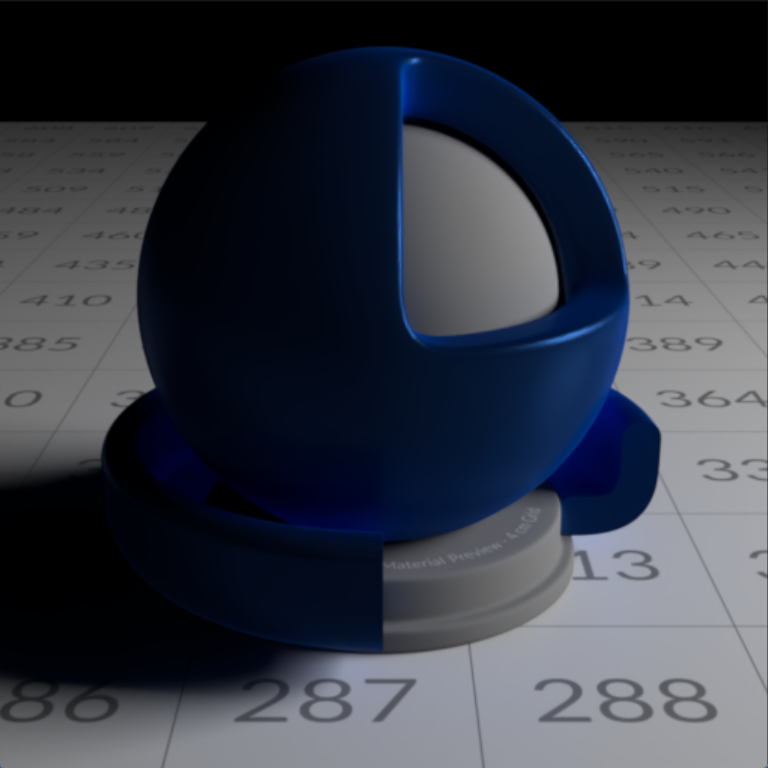}};
    \node[inner sep=1pt,above = 2pt of GGX0] { \textbf{$\Delta\theta_{\vecv\vecl}=0^\circ$} };
    \node[inner sep=1pt,above = 2pt of GGX45] { \textbf{$\Delta\theta_{\vecv\vecl}=45^\circ$} };
    \node[inner sep=1pt,above = 2pt of GGX90] { \textbf{$\Delta\theta_{\vecv\vecl}=90^\circ$} };
    \node[inner sep=1pt,left = 1pt of GGX0, anchor=south,rotate=90] { GGX };
    \node[inner sep=1pt,left = 1pt of MRM0, anchor=south,rotate=90] { MRM };
  \end{tikzpicture}
  \caption{Renderings of a metallic shaderball with either a classic GGX (top row) or retroreflective MRM material (bottom row) applied, for various camera angles and a fixed light direction, with relative angle between the view and light directions $\Delta\theta_{\vecv\vecl} \in \{0^\circ, 45^\circ, 90^\circ\}$.}
  \label{fig:conductor_renders}
  \vspace*{-6pt}
\end{figure}

Figure~\ref{fig:conductor_renders} shows renderings of a shaderball with either a classic GGX or retroreflective MRM material applied, here assuming \emph{conductive} (i.e.,\ metallic) microfacets.

\begin{figure}
  \centering
  \begin{tikzpicture}[font=\scriptsize ]
    \node[inner sep=1pt] (Class1) {\includegraphics[width=0.3\linewidth,frame]{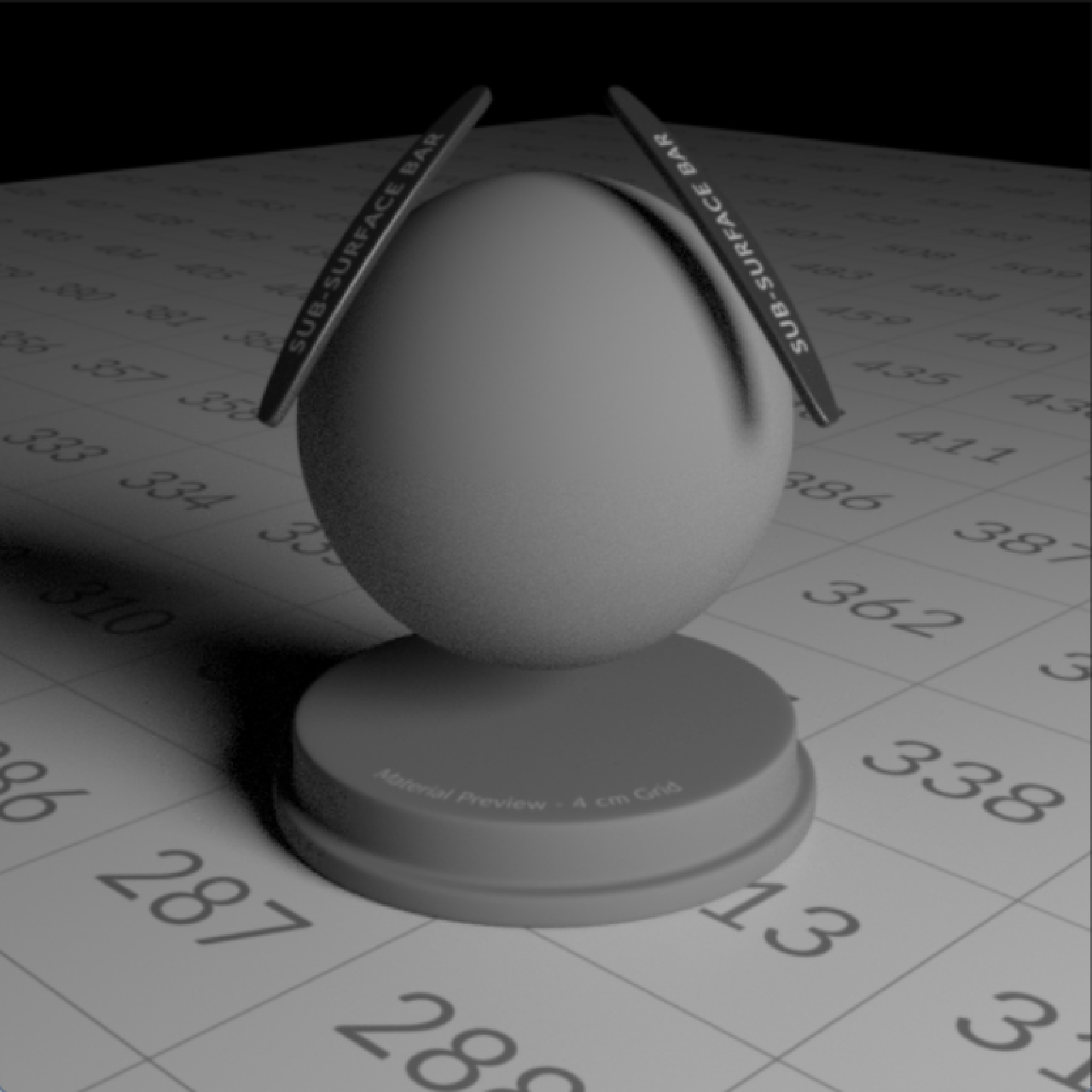 }};
    \node[inner sep=1pt,right = 0cm of Class1] (MRM1) {\includegraphics[width=0.3\linewidth,frame]{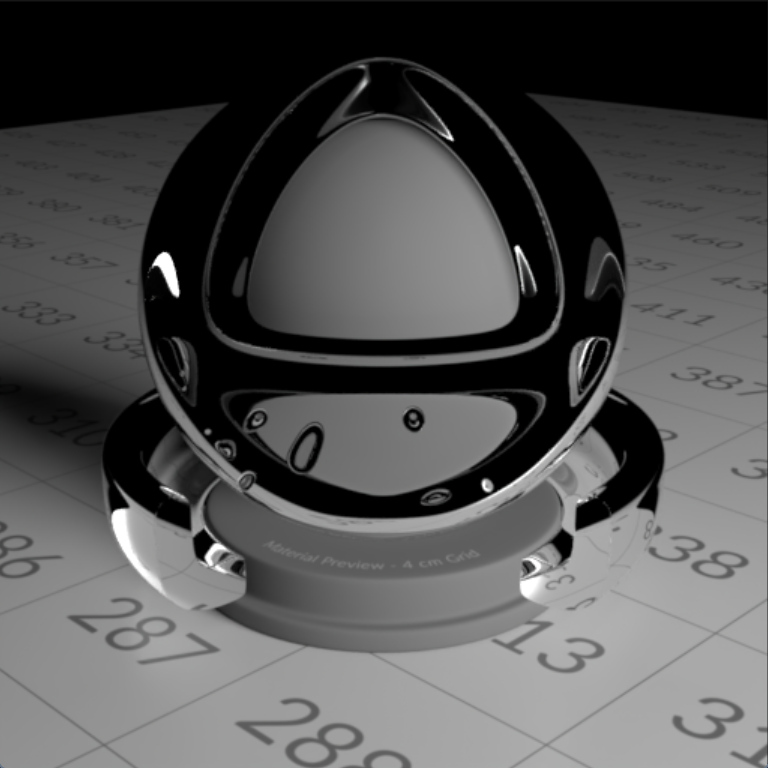 }};
    \node[inner sep=1pt,right = 0cm of MRM1] (Mixed1) {\includegraphics[width=0.3\linewidth,frame]{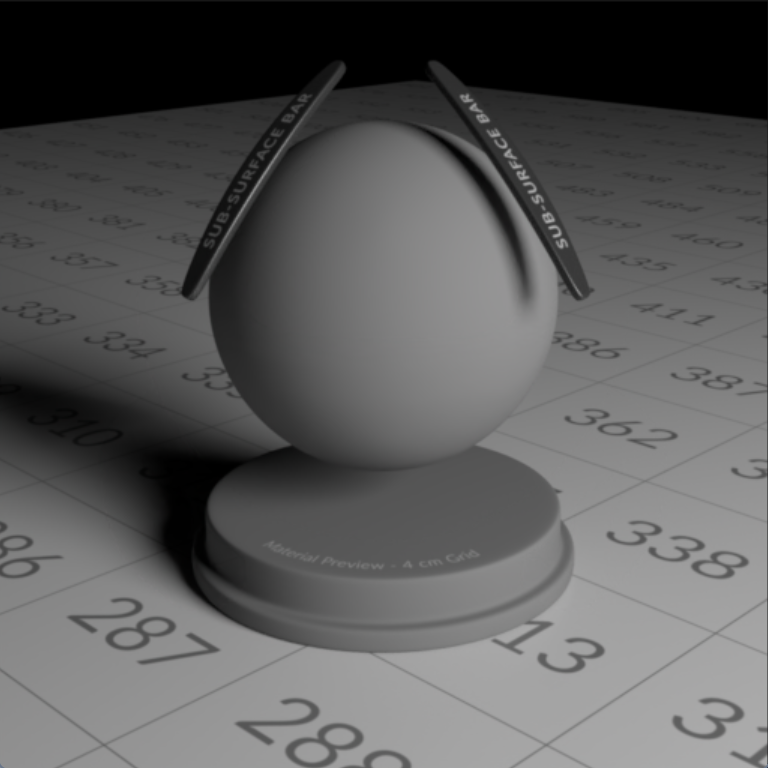 }};
    \node[inner sep=1pt,below = 0cm of Class1] (Class15) {\includegraphics[width=0.3\linewidth,frame]{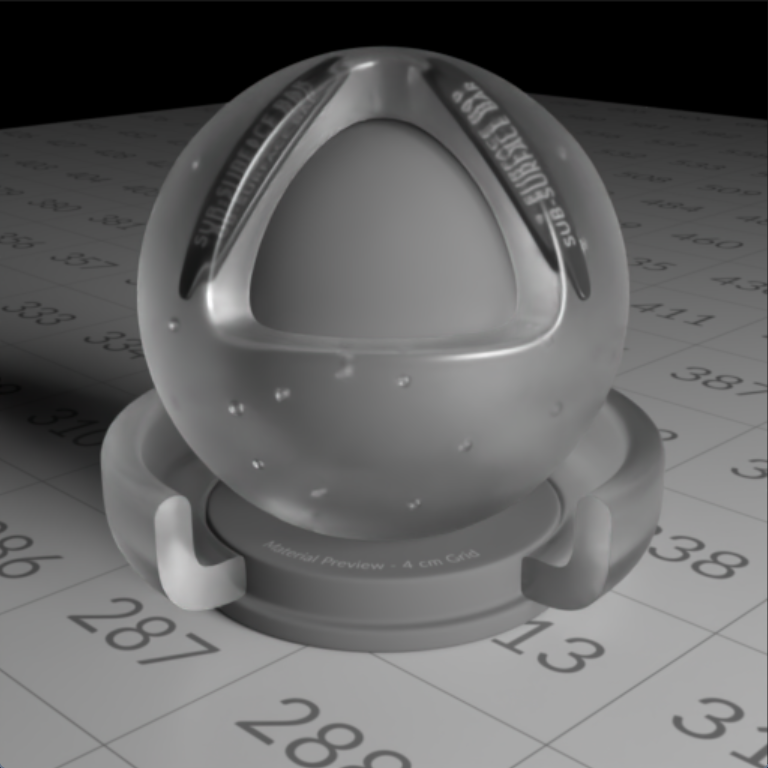 }};
    \node[inner sep=1pt,right = 0cm of Class15] (MRM15) {\includegraphics[width=0.3\linewidth,frame]{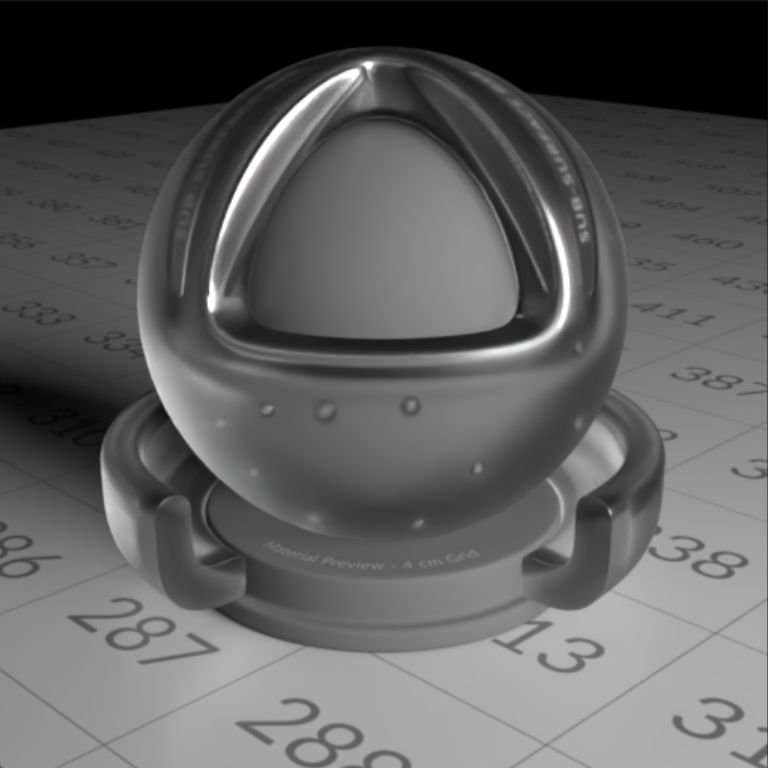 }};
    \node[inner sep=1pt,right = 0cm of MRM15] (Mixed15) {\includegraphics[width=0.3\linewidth,frame]{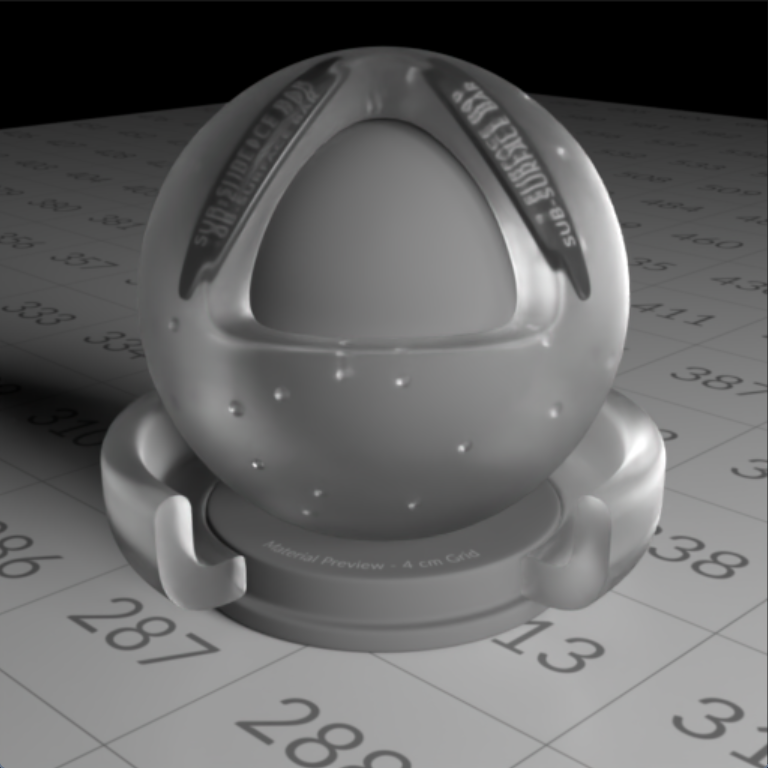 }};
    \node[inner sep=1pt,above = 2pt of Class1] { {Classic Microfacet} };
    \node[inner sep=1pt,above = 2pt of MRM1] { {Full Retroreflective MRM} };
    \node[inner sep=1pt,above = 11.5pt of Mixed1] { {Mixed MRM (BRDF)} };
    \node[inner sep=1pt,above = 0.5pt of Mixed1] { {and Classic (BTDF)} };
    \node[inner sep=1pt,left = 1pt of Class1, anchor=south,rotate=90] { $\mathbf{\boldsymbol{\eta} = 1}$ };
    \node[inner sep=1pt,left = 1pt of Class15, anchor=south,rotate=90] { $\mathbf{\boldsymbol{\eta} = 1.5}$ };
  \end{tikzpicture}
  \caption{Renderings of a glass shaderball illuminated from the side. The index of refraction is $1.0$ (top) or $1.5$ (bottom). From left to right: classic microfacet BSDF (left), retroreflective BSDF (middle), and combined retroreflective BRDF and classic BTDF (right).  }
  \label{fig:dielectric_renders}
\end{figure}

\paragraph*{Combination of BRDF and BTDF}

The albedo is identical for the forward- and retroreflective BRDF as well as the BTDF, which allows one to arbitrarily combine forward- and retroreflective variants into a BSDF. When constructing an artist-controllable retroreflective dielectric BSDF, we observed that using $f_t$ instead of $f_{t, \text{MRM}}$ yields more intuitive results, in particular for an index of refraction close to $1$. An implementation could reasonably choose either this formulation or the full retroreflective BTDF, depending on the desired effect.

Figure~\ref{fig:dielectric_renders} demonstrates this in renderings of a dielectric shaderball, where the IOR of the dielectric material is either $1.0$ (top row) or $1.5$ (bottom row).
From left to right, we show three combinations of BRDF and BTDF: the classic microfacet BSDF $f_r$ and BTDF $f_t$ (left), the retroreflective BRDF $f_{r,\text{MRM}}$ plus retroreflective BTDF $f_{t,\text{MRM}}$ (middle), and retroreflective BRDF $f_{r,\text{MRM}}$ combined with the classic (non-retroreflective) BTDF $f_t$ (right).

\section{Conclusion}

We have presented the Minimal Retroreflective Microfacet (MRM) model, which turns any existing microfacet BSDF into a retroreflective one by a single substitution: reflecting the view direction about the surface normal. We have shown that this modification preserves reciprocity and energy conservation, making MRM a physically plausible drop-in replacement. A comparison to measured retroreflective material data confirms that the model can represent materials with strong retroreflective properties---such as safety tape, certain fabrics, or retroreflective coatings---with reasonable accuracy and negligible implementation cost.

Due to its simplicity and compatibility with existing microfacet infrastructure, MRM has been adopted as part of the OpenPBR material standard \cite{OpenPBR2024} and the MaterialX shading framework \cite{Smythe2016}. Future work could explore fitting to a wider range of measured retroreflective materials and validating the retroreflective BTDF component against physical data.

\section*{Acknowledgements}
We thank Devin Miller for the shoe model in Figure~\ref{fig:teaser} and Masuo Suzuki for helpful discussions.

\small

\bibliographystyle{jcgt}
\bibliography{citations}

\section*{Author Contact Information}

\begin{tabular}{@{}p{0.49\textwidth}@{}p{0.49\textwidth}@{}}
Jamie Portsmouth \newline
\href{mailto:jamports@mac.com}{jamports@mac.com} \newline
&
Matthias Raab \newline
\href{mailto:mraab@nvidia.com}{mraab@nvidia.com} \newline
\\

Laurent Belcour \newline
\href{mailto:laurent.belcour@gmail.com}{laurent.belcour@gmail.com} 
&
Francis Liu \newline
\href{mailto:frankie.liu@gmail.com}{frankie.liu@gmail.com} 

\end{tabular}

\afterdoc

\end{document}